\documentclass[12pt,twoside, a4paper]{article}
\def\pd{\partial}
\def\mc{\mathcal}
\def\ul{\underline}

\usepackage[dvips]{graphicx}
\usepackage{amssymb}
\usepackage[bbgreekl]{mathbbol}
\usepackage{amssymb,amsmath}
\usepackage{graphicx}
\usepackage{dsfont}
\usepackage{caption}
\usepackage{subcaption}
\usepackage{mathtools}
\usepackage{verbatim}
\usepackage{graphicx}
\usepackage{multirow}
\usepackage[outline]{contour}
\usepackage{xcolor,colortbl}
\input{epsf.sty}  
\begin{document}
\begin{center}
\Large{\textbf{Twisted compactifications and conformal defects from ISO(3)$\times$U(1) $F(4)$ gauged supergravity}}
\end{center}
\vspace{1 cm}
\begin{center}
\large{\textbf{Parinya Karndumri}}
\end{center}
\begin{center}
String Theory and Supergravity Group, Department
of Physics, Faculty of Science, Chulalongkorn University, 254 Phayathai Road, Pathumwan, Bangkok 10330, Thailand \\
E-mail: parinya.ka@hotmail.com \vspace{1 cm}
\end{center}
\begin{abstract}
We study holographic solutions describing RG flows across dimensions from five-dimensional $N=2$ SCFT to SCFTs in three and two dimensions using matter-coupled $F(4)$ gauged supergravity with $ISO(3)\times U(1)$ gauge group. By performing topological twists implemented by $SO(2)$, $SO(2)\times SO(2)$ and $SO(3)$ gauge fields, we find a number of $AdS_4\times H^2$ and $AdS_3\times H^3$ solutions preserving eight and four supercharges. These geometries are identified with conformal fixed points in three and two dimensions, respectively. The corresponding flow solutions are obtained numerically and can be interpreted as black membranes and black strings in asymptotically $AdS_6$ space. In addition, we have constructed charged domain wall solutions with $AdS_2\times S^3$ and $AdS_3\times S^2$ slicing. These solutions are supported by a non-vanishing two-form field and describe conformal line and surface defects within the $N=2$ SCFT in five dimensions, respectively. All of these solutions can be uplifted to type IIB theory via consistent truncations on $S^2\times \Sigma$ with $\Sigma$ being a Riemann surface.
\end{abstract}
\newpage
%%%%%%%%%%%%%%%%%%%%%%%%%%%%%%%%%%%%%%%%%%%%%%%%%%%%%%%%%%%%%%%%%%%%%%%%%%%%%%%%%%%%%%%%%%%%%%%%%%%%%%%%%%%%%%%%%%%%%%%%%%%%%%%%%%%%%%%%%
\section{Introduction}
Five-dimensional field theories possess interesting non-trivial conformal fixed points at infinite coupling constants \cite{Seiberg_5Dfield,Seiberg_5Dfield3,Seiberg_5Dfield2}. These fixed points are interesting in their own right due to the fact that they realize the only possible $N=2$ superconformal symmetry in five dimensions \cite{Nahm_res}. The essentially strongly-coupled nature and the lack of a Lagrangian formulation make understanding these field theories and the corresponding fixed points a non-trivial task. Up to now, it is well-known that the AdS/CFT correspondence \cite{maldacena,Gubser_AdS_CFT,Witten_AdS_CFT} offers a new framework to gain more insights into the dynamics of strongly-coupled field theories via gravity duals. In AdS$_6$/CFT$_5$ correspondence, studies along this line have led to many interesting results, see for example \cite{ferrara_AdS6}-\cite{F4_defect}. These results, giving rise to various insights into different aspects of the five-dimensional superconformal field theories (SCFTs), have been obtained from both ten-dimensional string theories and effective six-dimensional gauged supergravities. Among these, holographic descriptions of twisted compactifications to lower-dimensional SCFTs \cite{6D_twist,F4_nunez,AdS6_BH_Minwoo1,AdS6_BH_Zaffaroni,AdS6_BH_Minwoo,AdS6_BH}, as initiated in \cite{MN_nogo}, and different types of conformal defects \cite{AdS2_F4_Dibitetto,6D_surface_Dibitetto,line_defect_F4_Gutperle,Christoph1,Christoph2,Christoph3,Nicolo_6D_defect1,
Nicolo_6D_defect2,Nicolo_6D_defect3,F4_defect} have recently attracted much attention.  
\\
\indent In this paper, we are interested in solutions describing twisted compactifications of five-dimensional $N=2$ SCFT to three and two dimensions and conformal line and surface defects in the framework of six-dimensional $N=(1,1)$ gauged supergravity, usually called $F(4)$ gauged supergravity. All the previous works on these types of solutions have considered only pure $F(4)$ gauged supergravity \cite{F4_Romans} with $SU(2)$ gauge group or a simple extension to matter-coupled theory with $SU(2)\times SU(2)$ gauge group \cite{F4SUGRA1,F4SUGRA2}. The former can be embedded in massive type IIA theory using the result of \cite{F4_form_mIIA} while the higher-dimensional origin of the latter is currently unknown. This renders many interesting holographic solutions, including a non-trivial $AdS_6$ vacuum with $SU(2)_{\textrm{diag}}$ symmetry identified in \cite{F4_flow}, not fully useful in the holographic context due to the lack of embedding in string/M-theory. We will partially fill this gap by considering $F(4)$ gauged supergravity coupled to four vector multiplets with $ISO(3)\times U(1)$ gauge group. This gauged supergravity and a truncation to three vector multiplets with $ISO(3)$ gauge group have been shown in \cite{Henning_Malek_AdS7_6} to arise from consistent truncations of type IIB theory on $S^2\times \Sigma$ for a Riemann surface $\Sigma$. 
\\
\indent The study of $ISO(3)\times U(1)$ $F(4)$ gauged supergravity has been initiated in \cite{ISO3_flow} in which the supersymmetric $AdS_6$ vacuum preserving $SO(3)\subset ISO(3)$ symmetry and a number of holographic RG flows and Janus interfaces have been found. These solutions only involve the metric and scalars. In this paper, we extend these results to more complicated solutions with non-vanishing gauge fields or a running two-form field. We will first look at supersymmetric $AdS_4\times \Sigma$ and $AdS_3\times \mc{M}_3$ solutions in which $\Sigma$ and $\mc{M}_3$ are a Riemann surface and a $3$-manifold with constant curvature, respectively. We will also find solutions interpolating between these geometries and the aforementioned supersymmetric $AdS_6$ vacuum. These will describe RG flows across dimensions from five-dimensional $N=2$ SCFT to conformal fixed points in three and two dimensions dual to $AdS_4$ and $AdS_3$ geometries in the IR. 
\\
\indent For solutions describing conformal defects within $N=2$ SCFT in five dimensions, we will study solutions dual to line and surface defects preserving conformal symmetry in one and two dimensions. These solutions will be obtained by taking the metric ansatz in the form of $AdS_2\times S^3$- and $AdS_3\times S^2$-sliced domain walls. As we will see, the results found here and in \cite{ISO3_flow} indicate that many aspects of the solutions are very similar to those previously found in compact $SU(2)$ and $SU(2)\times SU(2)$ gauge groups \cite{F4_defect}. Although the structure of solutions is not rich as in the case of compact $SU(2)\times SU(2)$ gauge group due to the lack of a non-trivial $AdS_6$ vacuum in addition to the trivial one at the origin of the scalar manifold, many properties and even the form of the solutions are very similar. This is in agreement with the general expectation that solutions of gauged supergravity that has no known higher-dimensional origins might still capture some insights into the general aspect of strongly-coupled dynamics of the dual field theories.
\\
\indent The paper is organized as follows. In section \ref{6D_SO4gaugedN2}, we review the general structure of matter-coupled $F(4)$ gauged supergravity as constructed in \cite{F4SUGRA1, F4SUGRA2}. We then consider the case of four vector multiplets and $ISO(3)\times U(1)$ gauge group and present the supersymmetric $AdS_6$ vacuum studied in \cite{ISO3_flow}. In section \ref{twisted}, we study supersymmetric $AdS_4\times \Sigma$ and $AdS_3\times \mc{M}_3$ solutions using the domain wall ansatz with $\mathbb{R}^{1,2}\times \Sigma$ and $\mathbb{R}^{1,1}\times \mc{M}_3$ slicing and implementing topological twists by turning on $SO(2)\times U(1)$ and $SO(3)$ gauge fields. A number of $AdS_4\times H^2$ and $AdS_3\times H^3$ backgrounds including numerical solutions interpolating between these geometries and the supersymmetric $AdS_6$ vacuum will also be given. We then move to the study of solutions describing conformal line and surface defects in section \ref{defects}. In this case, the metric ansatz takes the forms of $AdS_2\times S^3$- and $AdS_3\times S^2$-sliced domain walls, respectively. Unlike the case of twisted compactifications in which the two-form field from the supergravity multiplet can be consistently truncated out, we consider solutions with a non-vanishing two-form field and set all the gauge fields to zero. We find solutions describing line and surface defects due to a source of a dimension-$4$ operator dual to the two-form field. Finally, we give some conclusions and comments in section \ref{conclusion}.  
%%%%%%%%%%%%%%%%%%%%%%%%%%%%%%%%%%%%%%%%%%%%%%%%%%%%%%%%%%%%%%%%%%%%%%%%%%%%%%%%%%%%%%%%%%%%%%%%%%%%%%%%%%%%%%%%%%%%%%%%%%%%%%%%%%%%%%%%%
\section{Matter-coupled $F(4)$ gauged supergravity with $ISO(3)\times U(1)$ gauge group}\label{6D_SO4gaugedN2}
In this section, we give a brief review of matter-coupled $F(4)$ gauged supergravity in six dimensions. The field content of the supergravity coupled to $n$ vector multiplets is given by 
\begin{equation}
\left(e^{\hat{\mu}}_\mu,\psi^A_\mu, A^\Lambda_\mu, B_{\mu\nu}, \chi^A,\lambda^I_A,\phi^{I\alpha}\right) .
\end{equation}
We use the convention that the metric signature is $(-+++++)$ while space-time and tangent space indices are denoted respectively by $\mu,\nu=0,\ldots ,5$ and $\hat{\mu},\hat{\nu}=0,\ldots, 5$. The bosonic fields are the graviton $e^{\hat{\mu}}_\mu$, a two-form field $B_{\mu\nu}$, $4+n$ vector fields $A^\Lambda=(A^\alpha_\mu,A^I)$, the dilaton $\sigma$ and $4n$ scalars $\phi^{\alpha I}$ from the $n$ vector multiplets. The fermionic fields are given by two gravitini $\psi^A_\mu$, two spin-$\frac{1}{2}$ fields $\chi^A$ and $2n$ gaugini $\lambda^I_A$. All spinors are Sympletic-Majorana-Weyl. For various types of indices, we mostly follow the convention of \cite{F4SUGRA1,F4SUGRA2} with $\alpha=(0,r)=0,1,2,3$ and $I=1,2,\ldots, n$. Indices $A,B,\ldots =1,2$ denote the fundamental representation of $SU(2)_R\sim USp(2)_R\sim SO(3)_R$ R-symmetry while the adjoint indices are given by $r,s,\ldots =1,2,3$.
\\
\indent The dilaton and $4n$ scalars from the vector multiplets are described by $\mathbb{R}^+\times SO(4,n)/SO(4)\times SO(n)$ coset with $\mathbb{R}^+$ corresponding to the dilaton. The $SO(4,n)/SO(4)\times SO(n)$ can be parametrized by a coset
representative ${L^\Lambda}_{\ul{\Sigma}}$ transforming under the global $SO(4,n)$ and local $SO(4)\times SO(n)$ symmetries by left and right multiplications respectively with indices $\Lambda,\ul{\Sigma}=0,\ldots , n+3$. The local index can be decomposed into $\ul{\Sigma}=(\alpha,I)=(0,r,I)$ leading to various components of the coset representative 
\begin{equation}
{L^\Lambda}_{\ul{\Sigma}}=({L^\Lambda}_\alpha,{L^\Lambda}_I).
\end{equation}
The inverse of ${L^\Lambda}_{\ul{\Sigma}}$ will be denoted by ${(L^{-1})^{\ul{\Lambda}}}_\Sigma=({(L^{-1})^{\alpha}}_\Sigma,{(L^{-1})^{I}}_\Sigma)$. Furthermore, $SO(4,n)$ indices are raised and lowered by the invariant tensor
\begin{equation}
\eta_{\Lambda\Sigma}=\eta^{\Lambda\Sigma}=(\delta_{\alpha\beta},-\delta_{IJ}).
\end{equation}
\indent The bosonic Lagrangian of the matter-coupled gauged supergravity can be written as
\begin{eqnarray}
e^{-1}\mathcal{L}&=&\frac{1}{4}R-\pd_\mu \sigma\pd^\mu \sigma
-\frac{1}{4}P^{I\alpha}_\mu P^{\mu}_{I\alpha}-\frac{1}{8}e^{-2\sigma}\mc{N}_{\Lambda\Sigma}\widehat{F}^\Lambda_{\mu\nu}
\widehat{F}^{\Sigma\mu\nu}-\frac{3}{64}e^{4\sigma}H_{\mu\nu\rho}H^{\mu\nu\rho}\nonumber \\
& &-V-\frac{1}{64}e^{-1}\epsilon^{\mu\nu\rho\sigma\lambda\tau}B_{\mu\nu}\left(\eta_{\Lambda\Sigma}\widehat{F}^\Lambda_{\rho\sigma}
\widehat{F}^\Sigma_{\lambda\tau}+mB_{\rho\sigma}\widehat{F}^\Lambda_{\lambda\tau}\delta^{\Lambda 0}+\frac{1}{3}m^2B_{\rho\sigma}B_{\lambda\tau}\right)\nonumber \\
\label{Lar}
\end{eqnarray}
with $e=\sqrt{-g}$. The corresponding field strength tensors are defined by
\begin{eqnarray}
\widehat{F}^\Lambda=F^\Lambda-m\delta^{\Lambda 0}B, \qquad F^\Lambda=dA^\Lambda-\frac{1}{2}f^{\phantom{\Sigma}\Lambda}_{\Sigma\phantom{\Lambda}\Gamma} A^\Sigma\wedge A^\Gamma,\qquad H=dB,
\end{eqnarray}
where $f^{\phantom{\Sigma}\Lambda}_{\Sigma\phantom{\Lambda}\Gamma}$ are structure constants of the gauge group. It is also useful to note the convention on differential forms used in \cite{F4SUGRA1, F4SUGRA2}
\begin{equation}
F^\Lambda=F^\Lambda_{\mu\nu} dx^\mu \wedge dx^\nu\qquad \textrm{and}\qquad H=H_{\mu\nu\rho}dx^\mu \wedge dx^\nu\wedge dx^\rho
\end{equation}
which are different from the usual convention. In particular, in this convention, we have
\begin{eqnarray}
F_{\mu\nu}&=&\frac{1}{2}(\pd_\mu A_\nu-\pd_\nu A_\mu-f^{\phantom{\Sigma}\Lambda}_{\Sigma\phantom{\Lambda}\Gamma}A^\Sigma_\mu A^\Gamma_\nu)\nonumber \\
\textrm{and}\qquad H_{\mu\nu\rho}&=&\frac{1}{3}(\pd_\mu B_{\nu\rho}+\pd_\nu B_{\rho\mu}+\pd_\rho B_{\mu\nu})
\end{eqnarray}
\indent The vielbein on $SO(4,n)/SO(4)\times SO(n)$, $P^{I\alpha}_\mu=P^{I\alpha}_x\pd_\mu\phi^x$, $x=1,\ldots, 4n$, appearing in the scalar kinetic term, can be obtained from the left-invariant 1-form
\begin{equation}
{\Omega^{\ul{\Lambda}}}_{\ul{\Sigma}}=
{(L^{-1})^{\ul{\Lambda}}}_{\Pi}\nabla {L^\Pi}_{\ul{\Sigma}}
\qquad \textrm{with}\qquad \nabla
{L^\Lambda}_{\ul{\Sigma}}={dL^\Lambda}_{\ul{\Sigma}}
-f^{\phantom{\Gamma}\Lambda}_{\Gamma\phantom{\Lambda}\Pi}A^\Gamma
{L^\Pi}_{\ul{\Sigma}}
\end{equation}
via  
\begin{equation}
{P^I}_{\alpha}=({P^I}_{0},{P^I}_{r})=(\Omega^I_{\phantom{a}0},\Omega^I_{\phantom{a}r}).
\end{equation}
Other components of ${\Omega^{\ul{\Lambda}}}_{\ul{\Sigma}}$ are identified as the $SO(4)\times SO(n)$ composite connections $(\Omega^{rs},\Omega^{r0},\Omega^{IJ})$. 
\\
\indent The symmetric scalar matrix $\mc{N}_{\Lambda\Sigma}$, appearing in the vector kinetic term, is defined by
\begin{eqnarray}
\mc{N}_{\Lambda\Sigma}=L_{\Lambda\alpha}{(L^{-1})^\alpha}_\Sigma-L_{\Lambda I}{(L^{-1})^I}_\Sigma=(\eta L L^T\eta)_{\Lambda\Sigma}\, .
\end{eqnarray}
The scalar potential is given by
\begin{eqnarray}
V&=&-e^{2\sigma}\left[\frac{1}{36}A^2+\frac{1}{4}B^iB_i+\frac{1}{4}\left(C^I_{\phantom{s}t}C_{It}+4D^I_{\phantom{s}t}D_{It}\right)\right]
+m^2e^{-6\sigma}\mc{N}_{00}\nonumber \\
& &-me^{-2\sigma}\left[\frac{2}{3}AL_{00}-2B^iL_{0i}\right]
\end{eqnarray}
with various components of fermion-shift matrices defined by 
\begin{eqnarray}
A&=&\epsilon^{rst}K_{rst},\qquad B^i=\epsilon^{ijk}K_{jk0},\\
C^{\phantom{ts}t}_I&=&\epsilon^{trs}K_{rIs},\qquad D_{It}=K_{0It}\, .
\end{eqnarray}
The ``boosted'' structure constants are in turn given by
\begin{eqnarray}
K_{rs\alpha}&=&f^{\phantom{\Lambda}\Gamma}_{\Lambda\phantom{\Gamma}\Sigma}{L^\Lambda}_r(L^{-1})_{s\Gamma}{L^\Sigma}_\alpha,\nonumber \\
K_{\alpha It}&=&f^{\phantom{\Lambda}\Gamma}_{\Lambda\phantom{\Gamma}\Sigma}{L^\Lambda}_\alpha (L^{-1})_{I\Gamma}{L^\Sigma}_t
\end{eqnarray}
for $\alpha=(0,r)$. 
\\ 
\indent Finally, supersymmetry transformations of fermionic fields are given by
\begin{eqnarray}
\delta\psi_{\mu
A}&=&D_\mu\epsilon_A-\frac{1}{24}\left(Ae^\sigma+6me^{-3\sigma}(L^{-1})_{00}\right)\epsilon_{AB}\gamma_\mu\epsilon^B\nonumber
\\
& &-\frac{1}{8}
\left(B_te^\sigma-2me^{-3\sigma}(L^{-1})_{t0}\right)\gamma^7\sigma^t_{AB}\gamma_\mu\epsilon^B\nonumber \\
&
&+\frac{i}{16}e^{-\sigma}\left[\epsilon_{AB}(L^{-1})_{0\Lambda}\gamma_7+\sigma^r_{AB}(L^{-1})_{r\Lambda}\right]
F^\Lambda_{\nu\lambda}(\gamma_\mu^{\phantom{s}\nu\lambda}
-6\delta^\nu_\mu\gamma^\lambda)\epsilon^B\nonumber \\
& &+\frac{i}{32}e^{2\sigma}H_{\nu\lambda\rho}\gamma_7({\gamma_\mu}^{\nu\lambda\rho}-3\delta_\mu^\nu\gamma^{\lambda\rho})\epsilon_A,\label{delta_psi}\\
\delta\chi_A&=&\frac{1}{2}\gamma^\mu\pd_\mu\sigma\epsilon_{AB}\epsilon^B+\frac{1}{24}
\left[Ae^\sigma-18me^{-3\sigma}(L^{-1})_{00}\right]\epsilon_{AB}\epsilon^B\nonumber
\\
& &-\frac{1}{8}
\left[B_te^\sigma+6me^{-3\sigma}(L^{-1})_{t0}\right]\gamma^7\sigma^t_{AB}\epsilon^B\nonumber
\\
& &-\frac{i}{16}e^{-\sigma}\left[\sigma^r_{AB}(L^{-1})_{r\Lambda}-\epsilon_{AB}(L^{-1})_{0\Lambda}\gamma_7\right]F^\Lambda_{\mu\nu}\gamma^{\mu\nu}\epsilon^B\nonumber \\
& &-\frac{i}{32}e^{2\sigma}H_{\nu\lambda\rho}\gamma_7\gamma^{\nu\lambda\rho}\epsilon_A,
\label{delta_chi}\\
%\end{eqnarray}
%\begin{eqnarray}
\delta
\lambda^{I}_A&=&P^I_{ri}\gamma^\mu\pd_\mu\phi^i\sigma^{r}_{\phantom{s}AB}\epsilon^B+P^I_{0i}
\gamma^7\gamma^\mu\pd_\mu\phi^i\epsilon_{AB}\epsilon^B-\left(2i\gamma^7D^I_{\phantom{s}t}+C^I_{\phantom{s}t}\right)
e^\sigma\sigma^t_{AB}\epsilon^B \nonumber
\\
& &+2me^{-3\sigma}(L^{-1})^I_{\phantom{ss}0}
\gamma^7\epsilon_{AB}\epsilon^B-\frac{i}{2}e^{-\sigma}(L^{-1})^I_{\phantom{s}\Lambda}F^\Lambda_{\mu\nu}
\gamma^{\mu\nu}\epsilon_{A}\label{delta_lambda}
\end{eqnarray}
with ${\sigma^{rA}}_B$ being Pauli matrices. $SU(2)_R$ fundamental indices $A,B$ are raised and lowered by $\epsilon^{AB}$ and $\epsilon_{AB}$ with the convention $T^A=\epsilon^{AB}T_B$ and $T_A=T^B\epsilon_{BA}$. In this paper, we will use the values of $\epsilon^{12}=\epsilon_{12}=-1$. The covariant derivative of $\epsilon_A$ is defined by
\begin{equation}
D_\mu \epsilon_A=\pd_\mu
\epsilon_A+\frac{1}{4}\omega_\mu^{ab}\gamma_{ab}\epsilon_A+\frac{i}{2}\sigma^r_{AB}
\left[\frac{1}{2}\epsilon^{rst}\Omega_{\mu st}-i\gamma_7
\Omega_{\mu r0}\right]\epsilon^B\, .
\end{equation}
It is also useful to define the composite connection in the form
\begin{equation}
Q_{\mu AB}=\frac{i}{2}\sigma^r_{AB}
\left[\frac{1}{2}\epsilon^{rst}\Omega_{\mu st}-i\gamma_7
\Omega_{\mu r0}\right].
\end{equation}
\indent In this paper, we are interested in the case of $n=4$ vector multiplets leading to $\mathbb{R}^+\times SO(4,4)$ global symmetry. The gauge group of interest here is $ISO(3)\times U(1)$ with $ISO(3)$ embedded in $SO(4,3)\subset SO(4,4)$, and the $U(1)$ factor is the abelian gauge symmetry associated with the fourth vector multiplet as in the ungauged theory. The structure constants corresponding to the $ISO(3)\sim SO(3)\ltimes \mathbf{R}^3$ factor are given by
\begin{eqnarray}
f_{rst}=g\epsilon_{rst},\qquad f_{r\bar{s}\bar{t}}=-g\epsilon_{r\bar{s}\bar{t}},\qquad f_{\bar{r}\bar{s}\bar{t}}=-2g\epsilon_{\bar{r}\bar{s}\bar{t}}
\end{eqnarray}
with indices $I,J=1,2,3,4$ split as $I=(\bar{r},4)$ for $\bar{r},\bar{s}=1,2,3$. The resulting gauge generators are given by $X_\Lambda=(X_0,X_r,X_{\bar{r}})$ with $X_0=0$ while $X_r$ and $X_r-X_{\bar{r}}$ respectively generate the $SO(3)$ compact subgroup and three abelian translations $\mathbf{R}^3$ transforming as adjoint representation of $SO(3)$, see more detail in \cite{ISO3_flow}.
\\
\indent This gauged supergravity admits a supersymmetric $AdS_6$ vacuum at the origin of the $SO(4,4)/SO(4)\times SO(4)$ scalar manifold. This vacuum is given by, for more detail see \cite{ISO3_flow},
\begin{eqnarray}
& &{L^\Lambda}_{\ul{\Sigma}}={\delta^\Lambda}_{\Sigma},\qquad \sigma=\frac{1}{4}\ln\left[\frac{3m}{g}\right],\nonumber \\ 
& &V_0=-\frac{20g^2}{3}\sqrt{\frac{m}{3g}},\qquad L^2=-\frac{5}{V_0}=\frac{3}{4g^2}\sqrt{\frac{3g}{m}}\, .\label{AdS6_vacuum}
\end{eqnarray}
$V_0$ and $L$ denote the cosmological constant and the $AdS_6$ radius, respectively. As usual, we can choose $g=3m$ by shifting the value of the dilaton to $\sigma=0$ at the vacuum. Furthermore, this choice is also required by the embedding of six-dimensional gauged supergravity to type IIB theory given in \cite{Henning_Malek_AdS7_6}. With the value of the coupling constant $g=3m$, we find 
\begin{equation}
V_0=-20m^2\qquad \textrm{and}\qquad L=\frac{1}{2m}
\end{equation}
in which we have chosen $m>0$. This $AdS_6$ vacuum is dual to an $N=2$ SCFT in five dimensions.
%%%%%%%%%%%%%%%%%%%%%%%%%%%%%%%%%%%%%%%%%%%%%%%%%%%%%%%%%%%%%%%%%%%%%%%%%%%%%%%%%%%%%%%%%%%%%%%%%%%%%%%%%%%%%%%%%%%%%%%%%%%%%%%%%%%%%%%%%
\section{Twisted compactifications of $N=2$ SCFT from ISO(3)$\times$U(1) $F(4)$ gauged supergravity}\label{twisted}
In this section, we consider supersymmetric $AdS_4\times \Sigma$ and $AdS_3\times \mc{M}_3$ solutions with $\Sigma$ and $\mc{M}_3$ being a Riemann surface and a constant curvature $3$-manifold, respectively. These solutions are dual to conformal fixed points in three and two dimensions of the five-dimensional $N=2$ SCFT dual to the supersymmetric $AdS_6$ vacuum mentioned in the previous section. The solutions interpolating between these geometries and the $AdS_6$ vacuum are then interpreted as RG flows across dimensions from five dimensions to these fixed points in lower dimensions. In order to preserve some amount of supersymmetry, we need to perform a topological twist along $\Sigma$ and $\mc{M}_3$ by turning on suitable gauge fields to cancel spin connections on these manifolds. The lower-dimensional fixed points then arise from twisted compactifications of the $N=2$ SCFT in five dimensions on $\Sigma$ and $\mc{M}_3$. Moreover, the solutions can also be interpreted as black membranes and black strings in asymptotically $AdS_6$ space with near horizon geometries given by $AdS_4\times \Sigma$ and $AdS_3\times \mc{M}_3$. 

\subsection{Compactifications to three dimensions}
\begin{figure}
         \centering
               \begin{subfigure}[b]{0.4\textwidth}
                 \includegraphics[width=\textwidth]{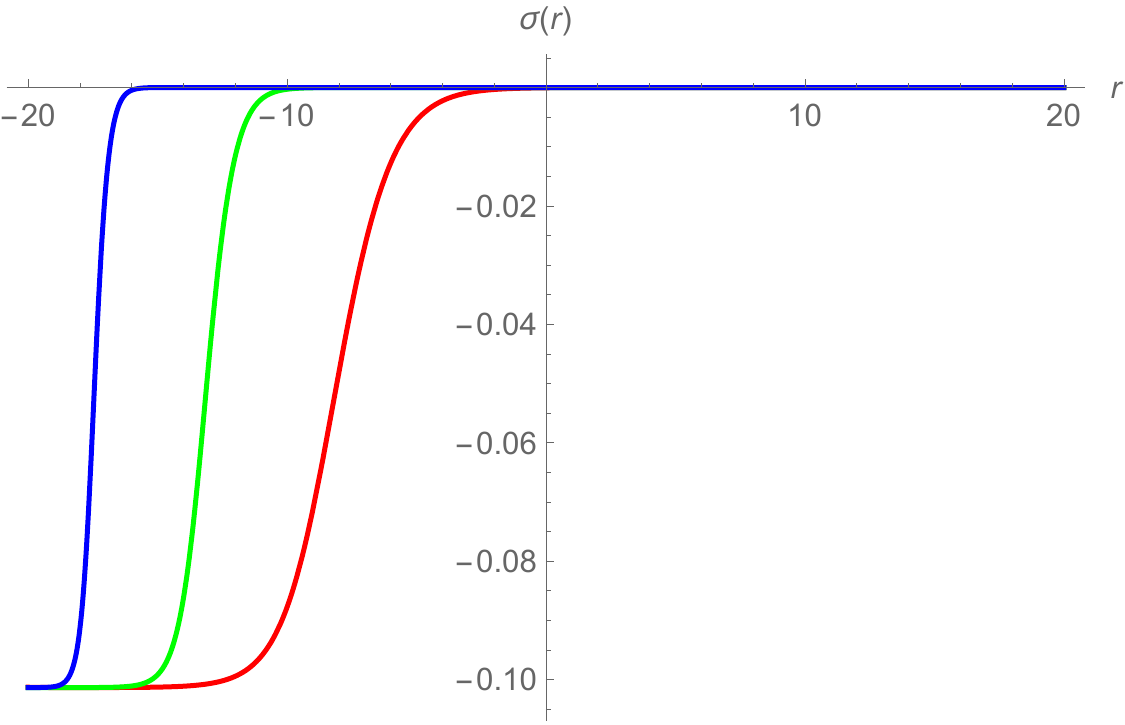}
                 \caption{Solutions for $\sigma(r)$}
         \end{subfigure}
         \begin{subfigure}[b]{0.4\textwidth}
                 \includegraphics[width=\textwidth]{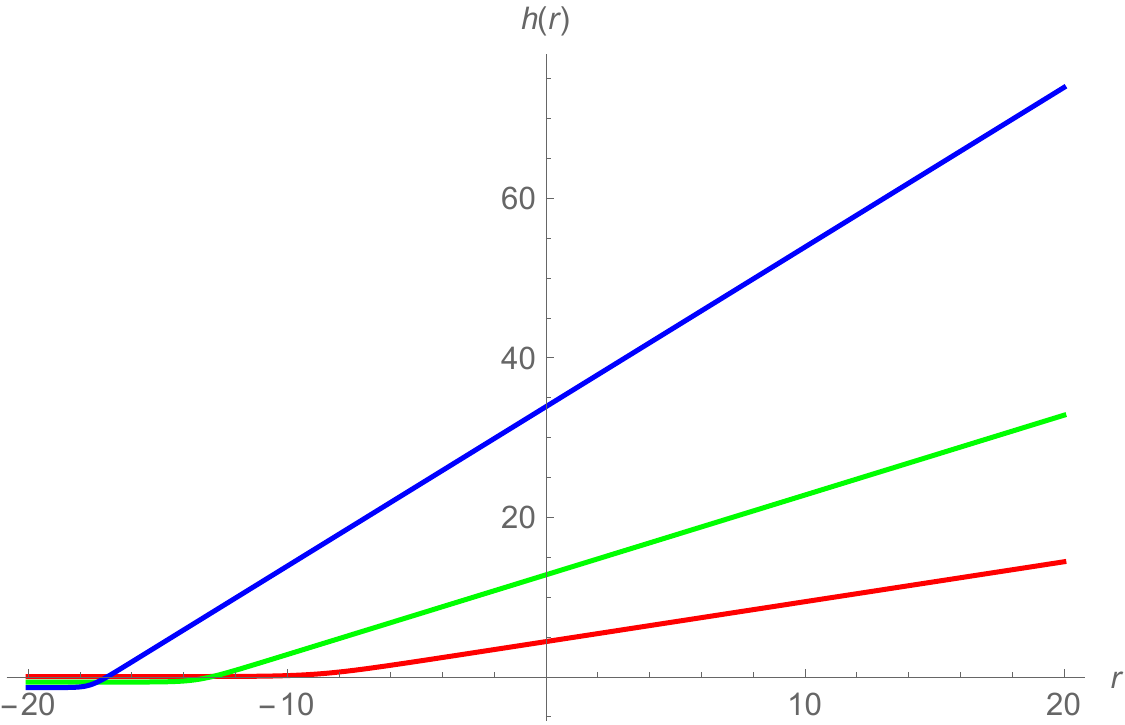}
                 \caption{Solutions for $h(r)$}
         \end{subfigure}
          \begin{subfigure}[b]{0.4\textwidth}
                 \includegraphics[width=\textwidth]{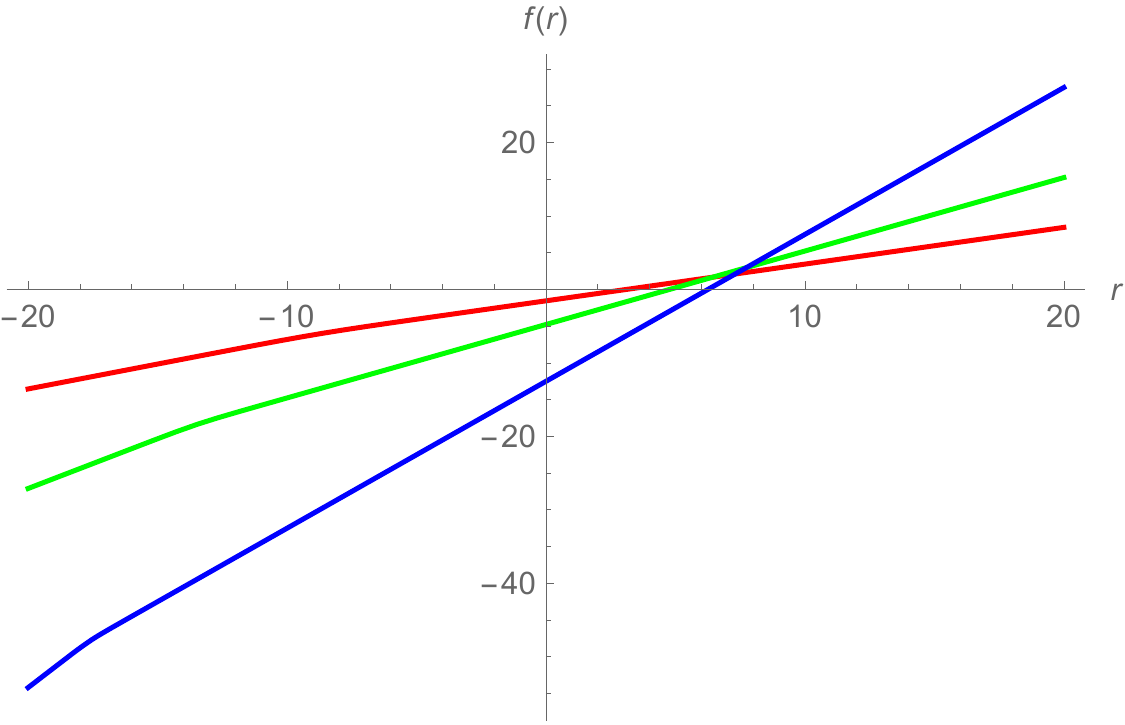}
                 \caption{Solutions for $f(r)$}
         \end{subfigure}
\caption{Supersymmetric RG flows from $N=2$ SCFT in five dimensions (dual to the $AdS_6$ vacuum \eqref{AdS6_vacuum}) to $N=2$ SCFT in three dimensions (dual to the $AdS_4\times H^2$ vacuum \eqref{AdS4_1}) from $ISO(3)$ $F(4)$ gauged supergravity with $g=3m$ and $m=\frac{1}{4}$ (red), $m=\frac{1}{2}$ (green), $m=1$ (blue).}\label{fig1}
 \end{figure}

We begin with twisted compactifications on a Riemann surface $\Sigma$. The metric ansatz is given by
\begin{equation}
ds^2=e^{2f(r)}dx_{1,2}^2+dr^2+e^{2h(r)}(d\theta^2+F_\kappa (\theta)^2d\phi^2)
\end{equation}
with $dx^2_{1,2}$ being a flat Minkowski metric in three dimensions with the signature $(-++)$. The function $F_\kappa(\theta)$ is defined by
\begin{equation}
F_\kappa(\theta)=\begin{cases}
  \sin\theta,  & \kappa=1\phantom{-}\quad \textrm{for}\quad \Sigma^2=S^2 \\
  \theta,  & \kappa=0\phantom{-}\quad \textrm{for}\quad \Sigma^2=T^2\\
  \sinh\theta,  & \kappa=-1\quad \textrm{for}\quad \Sigma^2=H^2
\end{cases}.\label{F_def}
\end{equation}
Denoting the six-dimensional coordinates by $x^\mu=(x^a,r,\theta,\phi)$, for $a=0,1,2$, we can choose the following choice of vielbein 
\begin{equation}
e^{\hat{a}}=e^fdx^a,\qquad e^{\hat{r}}=dr,\qquad e^{\hat{\theta}}=e^hd\theta,\qquad e^{\hat{\phi}}=e^hF_\kappa (\theta)d\phi\, .
\end{equation}
For convenience, we also present non-vanishing components of the spin connection
\begin{eqnarray}
& &{\omega^{\hat{a}}}_{\hat{r}}=f'e^{\hat{a}},\qquad {\omega^{\hat{\theta}}}_{\hat{r}}=h'e^{\hat{\theta}},\nonumber \\
& &{\omega^{\hat{\phi}}}_{\hat{r}}=h'e^{\hat{\phi}},\qquad {\omega^{\hat{\phi}}}_{\hat{\theta}}=\frac{F_\kappa'(\theta)}{F_\kappa(\theta)}e^{-h}e^{\hat{\phi}}\, .
\end{eqnarray}
In our convention, we will always denote $r$-derivatives by $'$ except $F'_\kappa(\theta)=\frac{dF_\kappa(\theta)}{d\theta}$. The component ${\omega^{\hat{\phi}}}_{\hat{\theta}}$ corresponds to an internal component along $\Sigma$ and needs to be cancelled by the topological twist.
\\
\indent We now consider an $SO(2)\times U(1)$ invariant sector with $SO(2)\subset SO(3)\subset ISO(3)$. As shown in \cite{ISO3_flow}, there are six $SO(2)\times U(1)$ singlet scalars from $SO(4,4)/SO(4)\times SO(4)$ coset with the coset representative given by
\begin{equation}
L=e^{\phi_0Y_{03}}e^{\phi_1(Y_{11}+Y_{22})}e^{\phi_2Y_{33}}e^{\phi_3(Y_{12}-Y_{21})}e^{\phi_4Y_{04}}e^{\phi_5Y_{34}}\, .\label{L_SO2}
\end{equation}
The $SO(4,4)$ non-compact generators can be written as
\begin{eqnarray}
Y_{\alpha I}=e^{\alpha,I+3}+e^{I+3,\alpha}
\end{eqnarray}  
for 
\begin{equation}
(e^{\Lambda \Sigma})_{\Gamma \Pi}=\delta^\Lambda_{
\Gamma}\delta^\Sigma_{\Pi},\qquad \Lambda, \Sigma,\Gamma,
\Pi=0,\ldots ,7\, .
\end{equation}
\indent We recall that the $SO(2)$ residual symmetry is generated by $X_3$, and the $U(1)$ is associated with the gauge field $A^7$. To perform a topological twist, we then turn on $SO(2)\times U(1)$ gauge fields of the form
\begin{equation}
A^3=aF'_\kappa(\theta)d\phi\qquad \textrm{and}\qquad A^7=bF'_\kappa(\theta)d\phi\, .\label{SO2_U1_gauge_field}
\end{equation}
The constants $a$ and $b$ correspond to the magnetic charges on $\Sigma$. With this ansatz for the gauge fields, it is consistent to set the two-form field to zero. A straightforward computation of the composite connection gives
\begin{equation}
Q_{\mu AB}=-\frac{i}{2}gA_\mu^3\sigma^3_{AB}\, .
\end{equation}
We also note that only $A^3$ appears in the composite connection since the gravitini are not charged under $A^7$ corresponding the $U(1)$ symmetry. The spin connection ${\omega^{\hat{\phi}}}_{\hat{\theta}}$ only appears in $\delta\psi_{A\hat{\phi}}$ variation in which the relevant terms are given by
\begin{equation}
0=\frac{1}{2}\frac{F'_\kappa(\theta)}{F_{\kappa}(\theta)}e^{-h}\gamma_{\hat{\phi}\hat{\theta}}\epsilon_A-\frac{i}{2}ga\frac{F'_\kappa(\theta)}{F_{\kappa}(\theta)}e^{-h}\sigma^3_{AB}\epsilon^B+\ldots\, .\label{AdS4_pretwist}
\end{equation}
The remaining terms involving $h'$, scalars and gauge fields take a similar form as in $\delta \psi_{A\hat{\theta}}$ variation. These will eventually lead to a BPS equation for $h(r)$ after the cancellation of the two terms given in \eqref{AdS4_pretwist}.  
\\
\indent We now perform a topological twist by imposing a projector
\begin{equation}
\gamma_{\hat{\theta}\hat{\phi}}\epsilon_A=-i\sigma^3_{AB}\epsilon^B\label{AdS4_proj}
\end{equation}
and a twist condition
\begin{equation}
ga=1\, .
\end{equation}
By using the relation $F''_\kappa(\theta)=-\kappa F_\kappa(\theta)$, we can write the gauge field strength tensors as
\begin{equation}
F^3=-\kappa a F_\kappa(\theta) d\theta\wedge d\phi\qquad \textrm{and}\qquad F^7=-\kappa b F_\kappa(\theta) d\theta\wedge d\phi\, .
\end{equation}
\indent With all these results, we are ready to analyze the BPS conditions arising from setting supersymmetry transformations of fermions to zero. With only the $(\theta\phi)$-component of the field strength tensors non-vanishing, the field equations of gauge fields require $\phi_3=0$. Furthermore, consistency of the BPS equations also require $\phi_0=\phi_4=0$. By imposing another projector of the form
\begin{equation}
\gamma_{\hat{r}}\epsilon_A=-\epsilon_A,\label{Gamma_r_pro}
\end{equation}
we find the following BPS equations
\begin{eqnarray}
\phi'_1&=&2ge^{\sigma+2\phi_1}\cosh\phi_5\sinh\phi_2,\\
\phi'_2&=&\frac{2ge^\sigma}{cosh\phi_5}\left[\cosh\phi_2(e^{2\phi_1}-1)-\sinh\phi_2\right]-\frac{1}{2}a\kappa e^{-2h-\sigma}\textrm{sech}\phi_5\sinh\phi_2,\,\,\,\\
\phi'_5&=&-2ge^\sigma\sinh\phi_5\left[\cosh\phi_2-(e^{2\phi_1}-1)\sinh\phi_5\right]\nonumber \\
& &+\frac{1}{2}\kappa e^{-2h-\sigma}(b\cosh\phi_5-a\cosh\phi_2\sinh\phi_5), \label{phi5_eq_tem}
\end{eqnarray}
\begin{eqnarray}
\sigma'&=&\frac{3}{2}me^{-3\sigma}+e^{-2h-\sigma}\left[\cosh\phi_2\cosh\phi_5(\kappa a-4ge^{2h+2\sigma}) \right.\nonumber \\
& &\left. +8ge^{2h+2\sigma+\phi_1}\sinh\phi_1\cosh\phi_5\sinh\phi_2-\kappa b\sinh\phi_5\right],\\
h'&=&\frac{1}{2}me^{-3\sigma}+\frac{1}{8}e^{-2h-\sigma}\left[\cosh\phi_2\cosh\phi_5(4ge^{2h+2\sigma}+3\kappa a) \right. \nonumber \\
& &\left. -8ge^{2h+2\sigma+\phi_1}\sinh\phi_1\cosh\phi_5\sinh\phi_2-3\kappa b \sinh\phi_5\right],\\
f'&=&\frac{1}{2}me^{-3\sigma}+\frac{1}{8}e^{-2h-\sigma}\left[\cosh\phi_2\cosh\phi_5(4ge^{2h+2\sigma}-\kappa a) \right. \nonumber \\
& &\left. -8ge^{2h+2\sigma+\phi_1}\sinh\phi_1\cosh\phi_5\sinh\phi_2+\kappa b \sinh\phi_5\right].
\end{eqnarray}
It can be readily verified that these equations are compatible with the corresponding field equations. The sign choice in \eqref{Gamma_r_pro} has been chosen such that the $AdS_6$ vacuum appears at $r\rightarrow \infty$. In addition, for $a=b=0$ which effectively sets all the gauge fields to zero, these equations reduce to those considered in \cite{ISO3_flow} for finding holographic RG flow solutions.
\\
\indent From the BPS equations, we find two $AdS_4\times \Sigma$ fixed points. The first one is given by
\begin{eqnarray}
\phi_1&=&\phi_2=\phi_5=0,\qquad b=0,\qquad \sigma=\frac{1}{4}\ln\left[\frac{2m}{g}\right],\nonumber \\
h&=&\ln\left[-\frac{a\kappa}{2\sqrt{2gm}}\right],\qquad L_{AdS_4}=\frac{1}{(2mg^3)^{\frac{1}{4}}}\, .\label{AdS4_1}
\end{eqnarray}
It should be pointed out that, from equation \eqref{phi5_eq_tem}, $AdS_4$ fixed points with $\phi_5=0$ are possible only for $b=0$ which implies $A^7=0$. Another possibility to satisfy equation \eqref{phi5_eq_tem} for $\phi_5=0$ is given by setting $\kappa=0$, but this does not lead to any $AdS_4$ fixed points. From equation \eqref{AdS4_1} and $g=3m$ with $m>0$, it is clearly seen that only $\kappa=-1$ is possible leading to an $AdS_4\times H^2$ solution. This solution can also be interpreted as an $AdS_4\times H^2$ solution of $F(4)$ gauged supergravity coupled to three vector multiplets with $ISO(3)$ gauge group since all the fields from the fourth vector multiplet have been truncated out. 
\\
\indent Another $AdS_4\times \Sigma$ solution is given by
\begin{eqnarray}
\phi_1&=&\phi_2=0,\qquad \phi_5=\frac{1}{2}\ln\left[\frac{\sqrt{a^2+8b^2}-3b}{a-b}\right],\nonumber \\
\sigma&=&\frac{1}{4}\ln\left[\frac{m(a+4b-\sqrt{a^2+8b^2})}{2gb}\sqrt{\frac{a-b}{\sqrt{a^2+8b^2}-3b}}\right],\nonumber \\
h&=&\frac{1}{2}\ln\left[-\frac{\kappa (a-b)}{2\sqrt{2}g}\sqrt{\frac{bg}{m}}\left(\frac{\sqrt{a+4b-\sqrt{a^2+8b^2}}}{a+2b-\sqrt{a^2+8b^2}}\right)\left(\frac{\sqrt{a^2+8b^2}-3b}{a-b}\right)^{\frac{1}{4}}\right],\nonumber \\
\frac{1}{L_{AdS_4}}&=&2m^{\frac{1}{4}}\left(\frac{2gb}{a+4b-\sqrt{a^2+8b^2}}\right)^{\frac{3}{4}}\left(\frac{\sqrt{a^2+8b^2}-3b}{a-b}\right)^{\frac{3}{8}}\, .\label{AdS4_2}
\end{eqnarray}
We have also chosen $a>b$ in these equations. As in the previous solution, this fixed point exists only for $\kappa=-1$ leading to another $AdS_4\times H^2$ solution. These solutions should be dual to conformal fixed points in three dimensions arising from twisted compactifications of $N=2$ SCFT in five dimensions, dual to the $AdS_6$ vacuum, on a two-dimensional hyperbolic space $H^2$. The solutions interpolating between these backgrounds then describe holographic RG flows from five to three dimensions. On the other hand, these solutions can also be regarded as black membranes with the near horizon geometry given by $AdS_4\times H^2$ in asymptotically $AdS_6$ space.
\\
\indent Before giving interpolating solutions between $AdS_6$ and $AdS_4\times H^2$ geometries, we note the unbroken supersymmetry preserved by the aforementioned solutions. At the $AdS_4$ fixed points with constant scalars, the $\gamma_r$ projector given in \eqref{Gamma_r_pro} is irrelevant. Accordingly, the $AdS_4\times H^2$ solutions are half-BPS, due to the projector \eqref{AdS4_proj}, and preserve eight supercharges corresponding to $N=2$ superconformal symmetry in three dimensions. On the other hand, the flow solutions interpolating between $AdS_6$ and $AdS_4\times H^2$ vacua preserve only four supercharges due to the two projectors given in \eqref{AdS4_proj} and \eqref{Gamma_r_pro}. 
\\
\indent Examples of numerical solutions interpolating between the supersymmetric $AdS_6$ vauum and the $AdS_4\times H^2$ solution \eqref{AdS4_1} are given in figure \ref{fig1}. In these solutions, we have set $g=3m$ and $m=\frac{1}{4}$ (red), $m=\frac{1}{2}$ (green), $m=1$ (blue). Along the flows, we have $\phi_1=\phi_2=\phi_5=0$. From the scalar masses given in \cite{ISO3_flow}, $\phi_1$ and $\phi_2$ scalars are dual to irrelevant operators of dimensions $\Delta=6,8$. The dilaton $\sigma$ is on the other hand dual to a relevant operator of dimension $\Delta=3$. Near the $AdS_6$ critical point, we find 
\begin{equation}
\sigma\sim e^{-\frac{3r}{L}}
\end{equation}
for $L=\frac{1}{2m}$. Therefore, the flow solutions shown in figure \ref{fig1} are driven by vacuum expectation values of an operator of dimension $3$.

\begin{figure}
         \centering
               \begin{subfigure}[b]{0.4\textwidth}
                 \includegraphics[width=\textwidth]{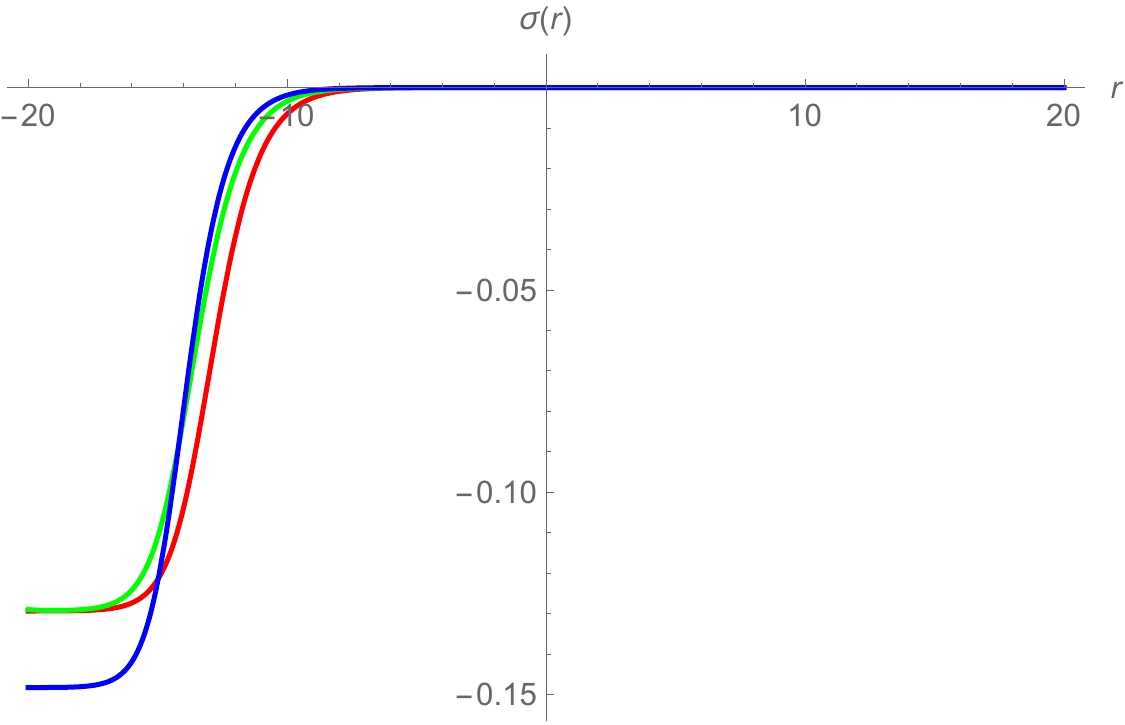}
                 \caption{Solutions for $\sigma(r)$}
         \end{subfigure}
         \begin{subfigure}[b]{0.4\textwidth}
                 \includegraphics[width=\textwidth]{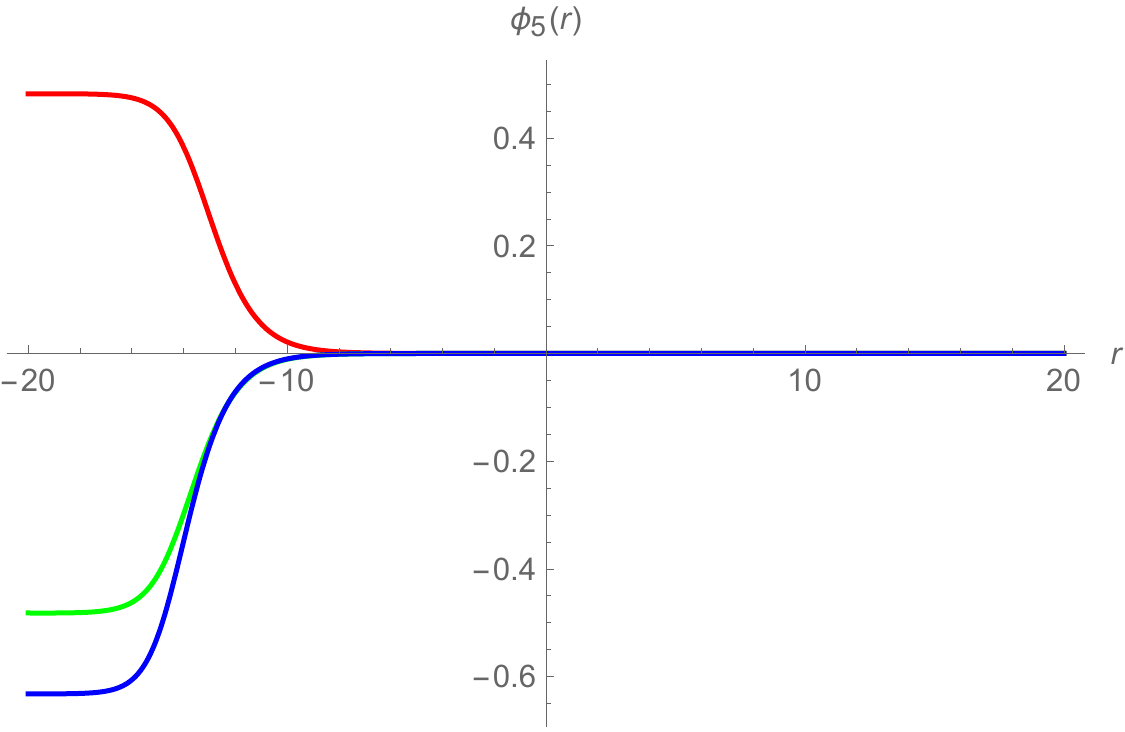}
                 \caption{Solutions for $\phi_5(r)$}
         \end{subfigure}\\
          \begin{subfigure}[b]{0.4\textwidth}
                 \includegraphics[width=\textwidth]{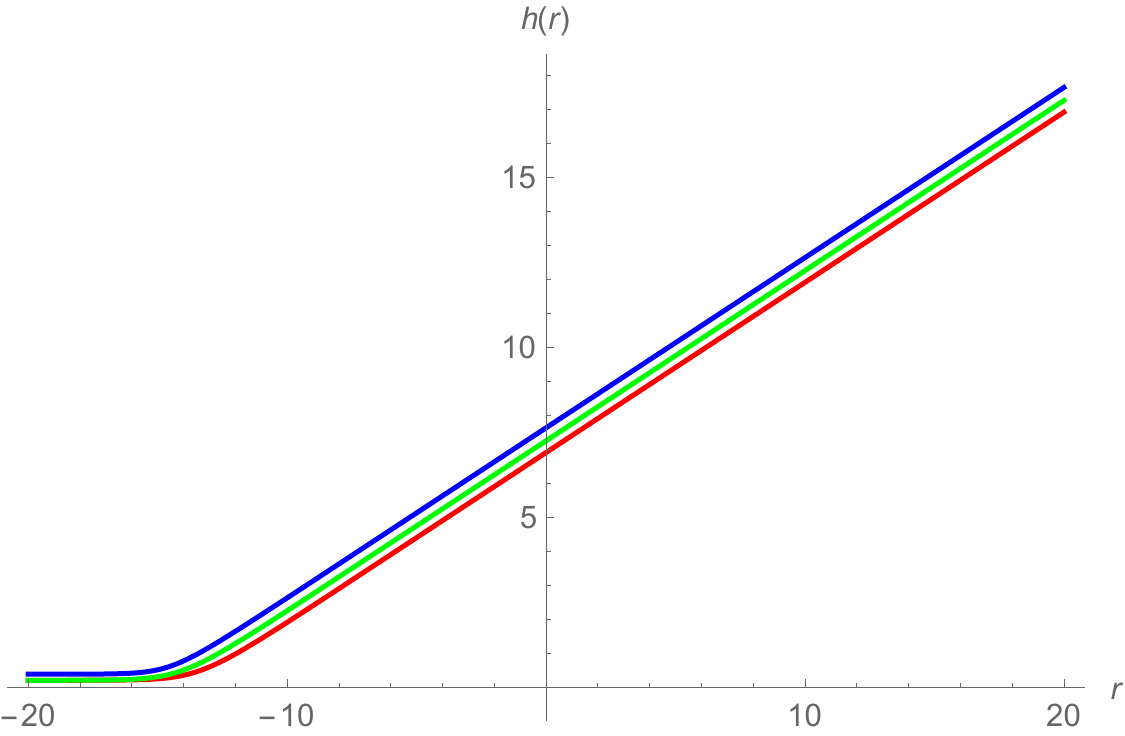}
                 \caption{Solutions for $h(r)$}
         \end{subfigure}
         \begin{subfigure}[b]{0.4\textwidth}
                 \includegraphics[width=\textwidth]{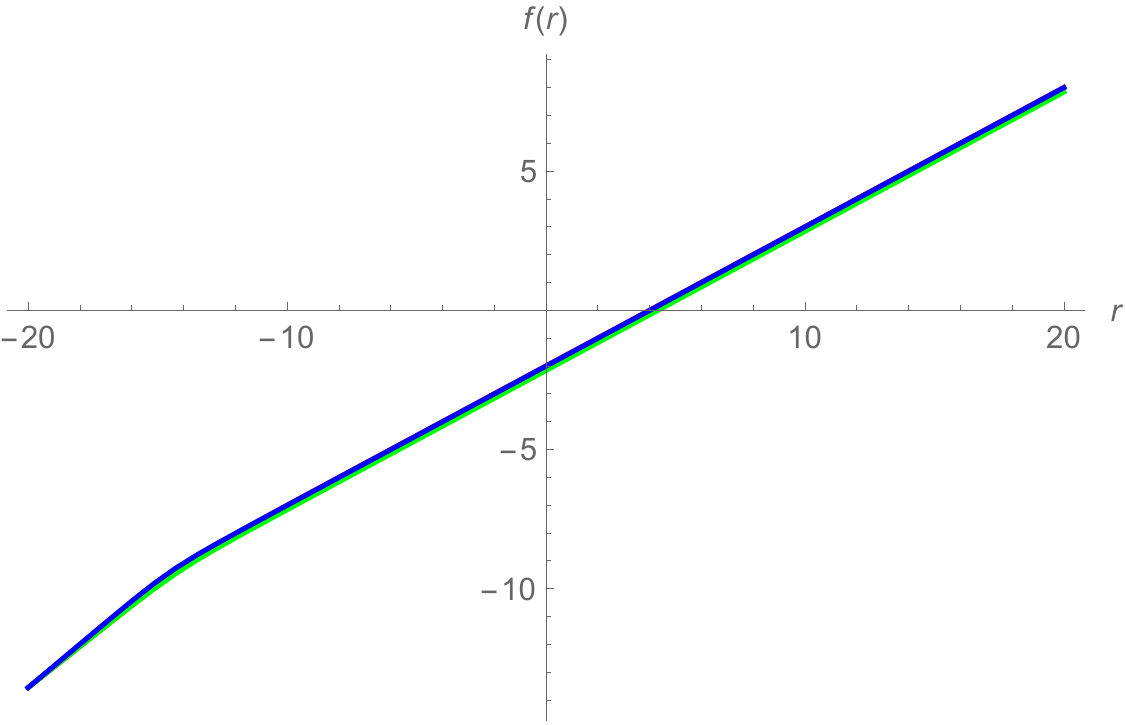}
                 \caption{Solutions for $f(r)$}
         \end{subfigure}
\caption{Supersymmetric RG flows from $N=2$ SCFT in five dimensions (dual to the $AdS_6$ vacuum \eqref{AdS6_vacuum}) to $N=2$ SCFT in three dimensions (dual to the $AdS_4\times H^2$ vacuum \eqref{AdS4_2}) from $ISO(3)\times U(1)$ $F(4)$ gauged supergravity with $g=3m$, $m=\frac{1}{4}$ and $b=-1$ (red), $b=1$ (green), $b=2$ (blue).}\label{fig2}
 \end{figure}

\indent Similarly, we can find examples of numerical flow solutions interpolating between the $AdS_6$ vacuum and $AdS_4\times H^2$ geometry \eqref{AdS4_2} as shown in figure \ref{fig2}. In these solutions, we have set $g=3m$, $m=\frac{1}{4}$ and $b=-1$ (red), $b=1$ (green), $b=2$ (blue). In this case, we have $\phi_1=\phi_2=0$ along the entire flows. As in the previous case, irrelevant operators are not turned on. In this case, near the $AdS_6$ critical point, we find  
\begin{equation}
\sigma\sim e^{-\frac{3r}{L}}\qquad \textrm{and}\qquad \phi_5\sim e^{-\frac{3r}{L}},
\end{equation}
so the flow solutions are driven by vacuum expectation values of two relevant operators of dimension $3$ dual to $\phi_5$ and $\sigma$.

\subsection{Compactifications to two dimensions}
\begin{figure}
         \centering
               \begin{subfigure}[b]{0.4\textwidth}
                 \includegraphics[width=\textwidth]{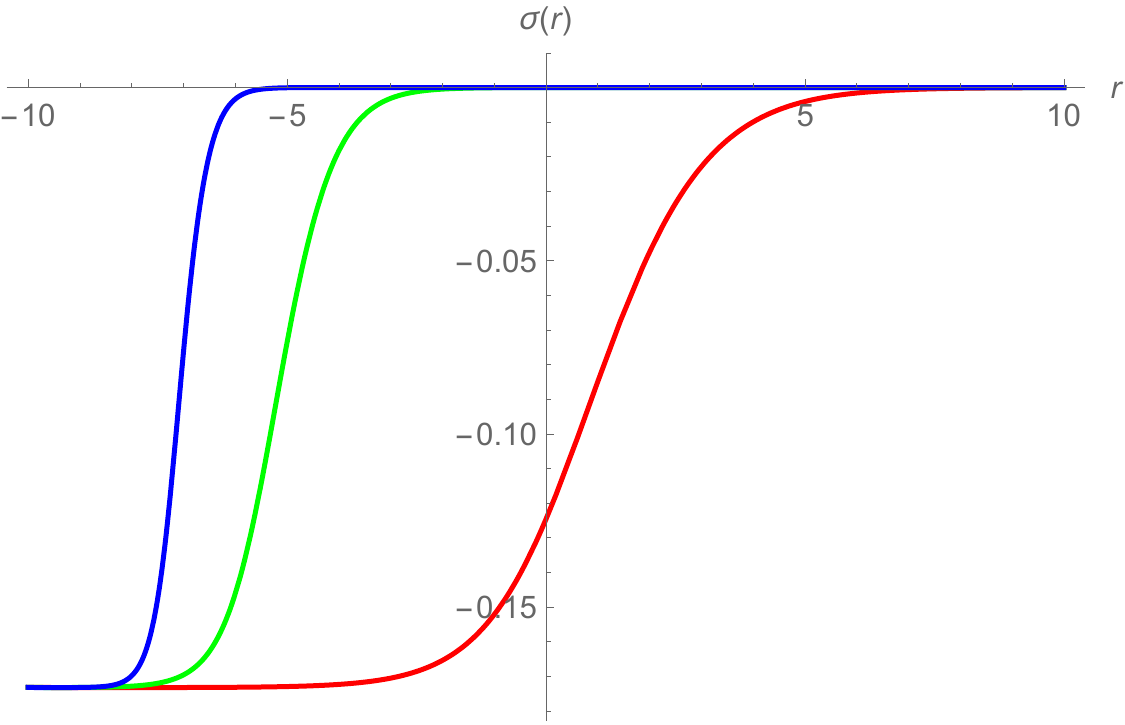}
                 \caption{Solutions for $\sigma(r)$}
         \end{subfigure}
         \begin{subfigure}[b]{0.4\textwidth}
                 \includegraphics[width=\textwidth]{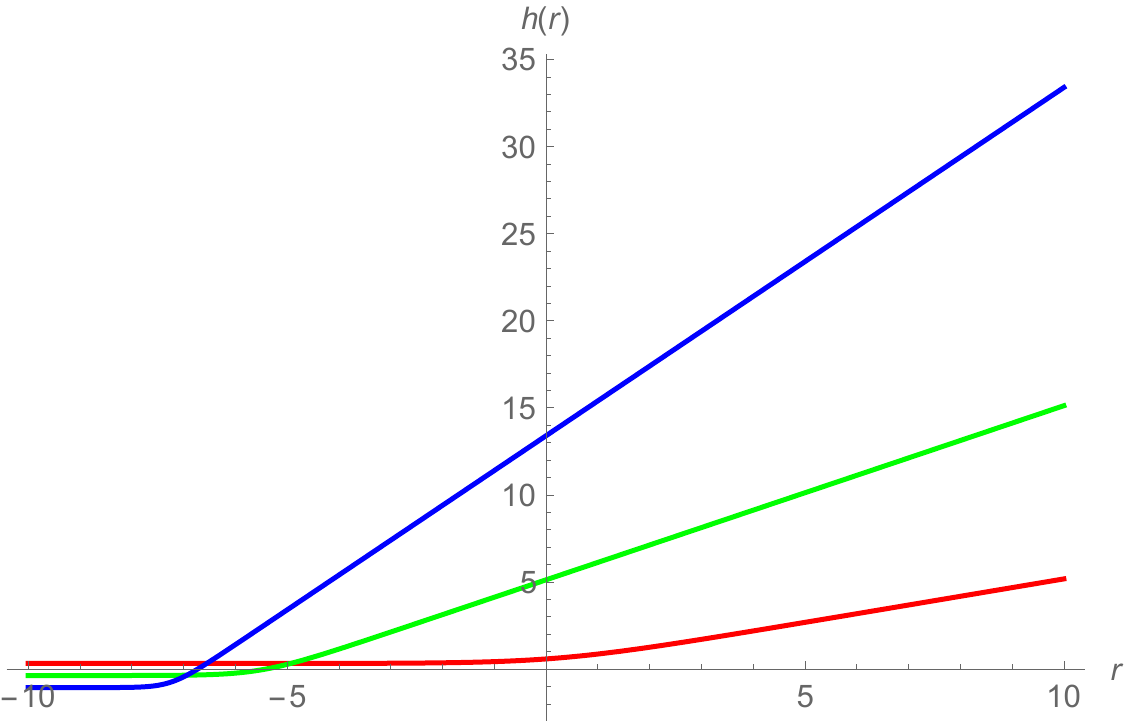}
                 \caption{Solutions for $h(r)$}
         \end{subfigure}
          \begin{subfigure}[b]{0.4\textwidth}
                 \includegraphics[width=\textwidth]{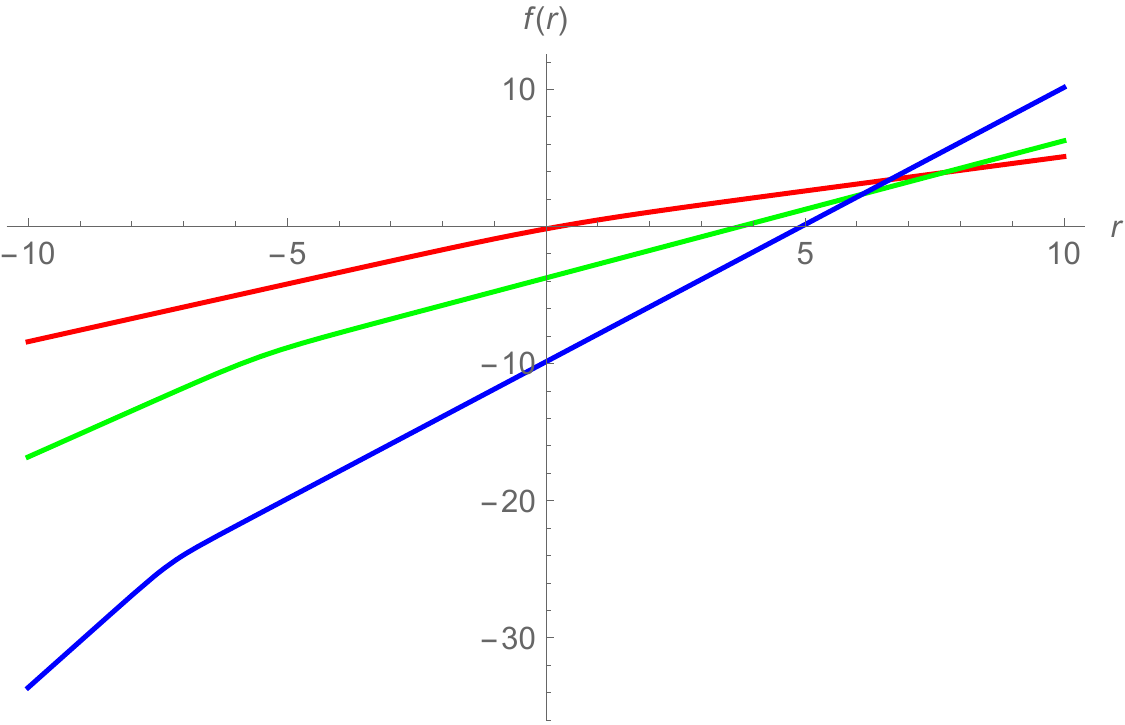}
                 \caption{Solutions for $f(r)$}
         \end{subfigure}
\caption{Supersymmetric RG flows from $N=2$ SCFT in five dimensions (dual to the $AdS_6$ vacuum \eqref{AdS6_vacuum}) to $N=(1,1)$ SCFT in two dimensions (dual to the $AdS_3\times H^3$ vacuum \eqref{AdS3_fixed_point}) from $ISO(3)$ $F(4)$ gauged supergravity with $g=3m$ and $m=\frac{1}{4}$ (red), $m=\frac{1}{2}$ (green), $m=1$ (blue).}\label{fig3}
 \end{figure}

In this section, we repeat a similar analysis for finding supersymmetric $AdS_3\times \mc{M}_3$ solutions to the $ISO(3)\times U(1)$ $F(4)$ gauged supergravity. We will work with a constant curvature manifold $\mc{M}_3$ described by the metric ansatz
\begin{equation}
ds^2=e^{2f(r)}dx^2_{1,1}+dr^2+e^{2h(r)}\left[d\psi^2+F_\kappa(\psi)^2(d\theta^2+\sin^2\theta d\phi^2)\right]
\end{equation}
with
\begin{equation}
F_\kappa(\psi)=\begin{cases}
  \sin\psi,  & \kappa=1\phantom{-}\quad \textrm{for}\quad \mc{M}_3=S^3 \\
  \psi,  & \kappa=0\phantom{-}\quad \textrm{for}\quad \mc{M}_3=T^3\\
  \sinh\psi,  & \kappa=-1\quad \textrm{for}\quad \mc{M}_3=H^3
\end{cases}\, .\label{F_def}
\end{equation}
In this case, $dx^2_{1,1}$ is the flat metric on two-dimensional Minkowski space with the signature $(-+)$.
\\
\indent Using the following choice of vielbein
\begin{eqnarray}
& &e^{\hat{a}}=e^fdx^a,\qquad e^{\hat{r}}=dr,\qquad e^{\hat{\psi}}=e^hd\psi,\nonumber \\
& &e^{\hat{\theta}}=e^hF_\kappa(\psi)d\theta,\qquad e^{\hat{\phi}}=e^hF_\kappa(\psi)\sin\theta d\phi,
\end{eqnarray}
with $a=0,1$, we find non-vanishing components of the spin connection of the form
\begin{eqnarray}
& &{\omega^{\hat{a}}}_{\hat{r}}=f'e^{\hat{a}},\qquad {\omega^{\hat{\psi}}}_{\hat{r}}=h'e^{\hat{\psi}},\qquad {\omega^{\hat{\theta}}}_{\hat{r}}=h'e^{\hat{\theta}},\qquad {\omega^{\hat{\phi}}}_{\hat{r}}=h'e^{\hat{\phi}},\nonumber \\
& &{\omega^{\hat{\phi}}}_{\hat{\psi}}=\frac{F'_\kappa(\psi)}{F_\kappa(\psi)}e^{-h}e^{\hat{\phi}},\qquad {\omega^{\hat{\phi}}}_{\hat{\theta}}=\frac{e^{-h}}{F_\kappa(\psi)}\cot \theta e^{\hat{\phi}},\qquad {\omega^{\hat{\theta}}}_{\hat{\psi}}=\frac{F'_\kappa(\psi)}{F_\kappa(\psi)}e^{-h}e^{\hat{\theta}}\, .\quad
\end{eqnarray}
We first note that the spin connection contains internal components on $\mc{M}_3$ only along $\theta$ and $\phi$ directions. To cancel these connections by performing a topological twist, we turn on the following $SO(3)\subset ISO(3)$ gauge fields
\begin{equation}
A^1=-a\cos\theta d\phi,\qquad A^2=-aF'_\kappa(\psi)\sin\theta d\phi,\qquad A^3=-aF'_\kappa(\psi) d\theta\, .\label{AdS3_gauge_field}
\end{equation}
We recall that this $SO(3)$ subgroup is generated by $X_r$ for $r=1,2,3$. There are two $SO(3)$ singlet scalars parametrized by the following coset representative, see more detail in \cite{ISO3_flow},
\begin{equation}
L=e^{\phi \hat{Y}_1}e^{\varphi \hat{Y}_2}\, .\label{L_SO3}
\end{equation} 
with
\begin{equation}
\hat{Y}_1=Y_{11}+Y_{22}+Y_{33}\qquad \textrm{and}\qquad \hat{Y}_2=Y_{04}\, .
\end{equation}
This leads to the composite connection of the form
\begin{equation}
Q_{\mu AB}=-\frac{i}{2}gA_\mu^r\sigma^r_{AB}\, .
\end{equation}
\indent The relevant terms in the gravitino variations are then given by
\begin{eqnarray}
\delta\psi_{A\hat{\theta}}:& &\, 0=\frac{1}{2}\frac{F'_\kappa(\psi)}{F_\kappa(\psi)}e^{-h}\gamma_{\hat{\theta}\hat{\psi}}\epsilon_A-\frac{i}{2}gA^r_{\hat{\theta}}\sigma^r_{AB}\epsilon^B+\ldots,\nonumber \\
\delta\psi_{A\hat{\phi}}:& &\, 0=\frac{1}{2}\frac{F'_\kappa(\psi)}{F_\kappa(\psi)}e^{-h}\gamma_{\hat{\phi}\hat{\psi}}\epsilon_A+\frac{1}{2}\frac{\cot\theta}{F_\kappa(\psi)}e^{-h}\gamma_{\hat{\phi}\hat{\theta}}\epsilon_A-\frac{i}{2}gA^r_{\hat{\phi}}\sigma^r_{AB}\epsilon^B+\ldots\quad\, .\label{AdS3_twist_eq}
\end{eqnarray}
The twist is achieved by imposing the twist condition
\begin{equation}
ag=1
\end{equation}
and the following projectors
\begin{equation}
\gamma_{\hat{\theta}\hat{\psi}}=-i\sigma^3_{AB}\epsilon^B,\qquad \gamma_{\hat{\phi}\hat{\psi}}=-i\sigma^2_{AB}\epsilon^B,\qquad \gamma_{\hat{\phi}\hat{\theta}}=-i\sigma^1_{AB}\epsilon^B\, .\label{AdS3_proj}
\end{equation}
With all these conditions together with the ansatz for the $SO(3)$ gauge fields given in \eqref{AdS3_gauge_field}, all the terms appearing in \eqref{AdS3_twist_eq} cancel among each other.   
\\
\indent It should be noted that only two of the projectors in \eqref{AdS3_proj} are independent, so the $AdS_3\times \mc{M}_3$ solutions will preserve four supercharges. In the dual five-dimensional $N=2$ SCFT, the $16$ supercharges transform under $SO(1,4)\times SO(3)_R$ as $(\mathbf{4},\mathbf{2})$ and $(\bar{\mathbf{4}},\mathbf{2})$. Under the decomposition to two dimensions, we have $SO(1,4)\rightarrow SO(1,1)\times SO(3)$ under which $\mathbf{4}\rightarrow \mathbf{2}_{1}+\mathbf{2}_{-1}$. The original $16$ supercharges then transform under $SO(1,1)\times SO(3)\times SO(3)_R$ as
\begin{eqnarray}
(\mathbf{4},\mathbf{2})+(\bar{\mathbf{4}},\mathbf{2})\rightarrow 2\times (\mathbf{2}_{1}+\mathbf{2}_{-1},\mathbf{2}). 
\end{eqnarray}
After imposing the topological twist by identifying the $SO(3)$ with the $SO(3)_R$ symmetry, we end up with the following decomposition
\begin{eqnarray}
(\mathbf{4},\mathbf{2})+(\bar{\mathbf{4}},\mathbf{2})\rightarrow 2\times (\mathbf{1}_{1}+\mathbf{1}_{-1}+\mathbf{3}_1+\mathbf{3}_{-1})
\end{eqnarray}
under $SO(1,1)\times SO(3)_{\textrm{diag}}$. The unbroken supersymmetry is then generated by the singlet supercharges and corresponds to $N=(1,1)$ superconformal symmetry in two dimensions, see also the discussion in \cite{F4_nunez}.
\\
\indent In order to derive the BPS equations, it is useful to note the explicit form of the gauge field strength tensors
\begin{equation}
F^1_{\hat{\theta}\hat{\phi}}=\frac{1}{2}\kappa ae^{-2h},\qquad F^2_{\hat{\psi}\hat{\phi}}=\frac{1}{2}\kappa ae^{-2h}, \qquad F^3_{\hat{\psi}\hat{\theta}}=\frac{1}{2}\kappa ae^{-2h}\, .
\end{equation} 
As in the previous case, with the ansatz for the gauge fields given above, it is consistent to set the two-form field to zero. With all these, we find the following BPS equations
\begin{eqnarray}
\phi'&=&\frac{1}{4}e^{-2h-\sigma-\phi}(e^{2\phi}-1)(4ge^{2h+2\sigma+2\phi}-\kappa a),\\
\varphi'&=&-2me^{-3\sigma}\sinh\varphi,\\
\sigma'&=&\frac{1}{4}e^{-3\sigma-\varphi}\left[ge^{4\sigma+\phi+\varphi}(e^{2\phi}-3)+3m(1+e^{2\varphi})\right]+\frac{3}{8}\kappa a e^{-2h-\sigma}\cosh\phi,\\
h'&=&\frac{1}{4}e^{-3\sigma}\left[ge^{4\sigma+\phi}(3-e^{2\phi})+m(e^\varphi+e^{-\varphi})\right]+\frac{5}{8}\kappa a e^{-2h-\sigma}\cosh\phi,\\
f'&=&\frac{1}{4}e^{-3\sigma}\left[ge^{4\sigma+\phi}(3-e^{2\phi})+m(e^\varphi+e^{-\varphi})\right]-\frac{3}{8}\kappa a e^{-2h-\sigma}\cosh\phi\, .
\end{eqnarray}
It can also be checked that all these equations satisfy the field equations. As in the previous case, for $a=0$, these equations reduce to those considered in \cite{ISO3_flow}. 
\\
\indent It turns out that these equations admit only one $AdS_3\times \mc{M}_3$ fixed point given by
\begin{eqnarray}
& &\phi=\varphi=0,\qquad \sigma= \frac{1}{4}\ln\left[\frac{3m}{2g}\right],\nonumber \\
& &h=\frac{1}{2}\ln\left[-\sqrt{\frac{3}{2}}\frac{\kappa a}{\sqrt{gm}}\right],\qquad \frac{1}{L_{AdS_3}}=2\left(\frac{2}{3}\right)^{\frac{3}{4}}m^{\frac{1}{4}}g^{\frac{3}{4}}\, .\label{AdS3_fixed_point}
\end{eqnarray}
The fixed point exists only for $\kappa=-1$ giving rise to an $AdS_3\times H^3$ solution. Examples of numerical solutions interpolating between this geometry and the $AdS_6$ vacuum are given in figure \ref{fig3} for $g=3m$ and $m=\frac{1}{4}$ (red), $m=\frac{1}{2}$ (green), $m=1$ (blue). Since $\varphi=0$ both at the $AdS_3\times H^3$ fixed point and along the flow, the solutions can also be regarded as solutions of $ISO(3)$ $F(4)$ gauged supergravity coupled to three vector multiplets. The solutions describe RG flows from the five-dimensional $N=2$ SCFT to a two-dimensional $N=(1,1)$ SCFT in the IR. Equivalently, the solutions can also be interpreted as supersymmetric black strings in asymptotically $AdS_6$ space. 
\\
\indent As in the previous case, $\sigma$ participates in the flows with the behavior near the $AdS_6$ vacuum 
\begin{equation}
\sigma\sim e^{-\frac{3r}{L}}\, .
\end{equation}
This again implies that the flow is driven by a vacuum expectation value of a dimension-$3$ operator.

%%%%%%%%%%%%%%%%%%%%%%%%%%%%%%%%%%%%%%%%%%%%%%%%%%%%%%%%%%%%%%%%%%%%%%%%%%%%%%%%%%%%%%%%%%%%%%%%%%%%%%%%%%%%%%%%%%%%%%%%%%%%%%%%%%%%%%%%%
\section{Line and surface defects from ISO(3)$\times$U(1) $F(4)$ gauged supergravity}\label{defects}
In this section, we consider another class of holographic solutions describing conformal line and surface defects within the $N=2$ SCFT in five dimensions. These solutions break five-dimensional superconformal symmetry but preserve conformal symmetries in one and two dimensions. The supergravity solutions realizing these defects take the form of curved domain walls with $AdS_2\times S^3$ and $AdS_3\times S^2$ slices, respectively. These curved domain walls are supported by a non-vanishing two-form field from the supergravity multiplet. For simplicity, we will truncate out all the vector fields. Therefore, the solutions we will consider only involve the metric, scalars and the two-form field.

\subsection{Line defects}
We first consider solutions that are holographically dual to conformal line defects within the $N=2$ SCFT in five dimensions. The metric ansatz takes the form
\begin{equation}
ds^2=e^{2f(r)}ds^2_{AdS_2}+dr^2+e^{2h(r)}ds^2_{S^3}\, .\label{line_metric}
\end{equation}
We will accordingly split the six-dimensional coordinates as $x^\mu =(x^a,r,\theta^i)$ for $a=0,1$ and $i=1,2,3$. As in \cite{AdS2_F4_Dibitetto} and \cite{F4_defect}, we will take the ansatz for the two-form field to be 
\begin{equation}
B=b(r)\textrm{vol}_{AdS_2}\, .
\end{equation}
As pointed out in \cite{6D_surface_Dibitetto}, with vanishing gauge fields, the action of supersymmetry transformations on $SU(2)_R$ indices $A,B$ is trivial. The two chiralities of the Symplectic-Majorana-Weyl spinors can then be conveniently combined into a Dirac spinor. The Killing spinors for the solutions of interest take the form
\begin{equation}
\epsilon=\epsilon^++\epsilon^-
\end{equation}
with
\begin{eqnarray}
& &\epsilon^+=iY(r)\left[\cos\theta(r)\chi^+\otimes \epsilon_0+\sin\theta(r)\chi^+\otimes \sigma_3\epsilon_0\right]\otimes \eta,\nonumber \\
& &\epsilon^-=Y(r)\left[\sin\theta(r)\chi^-\otimes \epsilon_0-\cos\theta(r)\chi^-\otimes \sigma_3\epsilon_0\right]\otimes \eta\, .
\end{eqnarray}  
In this equation, $\epsilon_0$ is a two-component constant spinor while $\chi^\pm$ and $\eta$ are respectively Killing spinors on $AdS_2$ and $S^3$ satisfying   
\begin{equation}
\hat{\nabla}_a\chi^\pm=\pm\frac{i}{2\ell }\rho_a\chi^\mp\qquad \textrm{and}\qquad \widetilde{\nabla}_i\eta=\frac{i}{2R}\tilde{\gamma}_i\eta\, .\label{AdS2_S3_Killing}
\end{equation}
$\ell$ and $R$ denote the radii of $AdS_2$ and $S^3$, respectively.  
\\
\indent As in \cite{F4_defect}, we use the following form of gamma matrices
\begin{eqnarray}
& &\gamma^a=\rho^a\otimes \sigma_2\otimes\mathbb{I}_2,\qquad \gamma^r=\rho_*\otimes \sigma_2\otimes \mathbb{I}_2,\nonumber \\
& &\gamma^i=-\mathbb{I}_2\otimes \sigma_1\otimes\tilde{\gamma}^i,\qquad \gamma_7=-i\mathbb{I}_2\otimes \sigma_3\otimes \mathbb{I}_2\, .
\end{eqnarray}
In these equations, the chirality matrix on $AdS_2$ is defined by $\rho_*=\rho^0\rho^1$ with $\rho^2_*=\mathbb{I}_2$.
\\
\indent We now analyze the BPS conditions arising from setting supersymmetry transformations of fermions to zero. The coset representative is taken to be the $SO(2)\times U(1)$ symmetric one given in \eqref{L_SO2}. Since we have set all the gauge fields to zero, we need to set $\phi_3=0$. Due to various similarities between the forms of relevant equations in this case and those of compact $SU(2)\times SU(2)$ gauge group considered \cite{F4_defect}, the analysis is essentially the same as in \cite{F4_defect} to which we refer for more detail. We first need to impose a projection condition on the Killing spinors
\begin{equation}
\sigma_2\epsilon_0=\epsilon_0\, .
\end{equation}
We also note that this is related to a $\gamma_r$ projector. We now consider the gaugino variations $\delta\lambda^I_A$. 
Since the two-form field does not appear in these variations, we simply obtain the following equations from $\delta\lambda^{1,2}_A$
\begin{eqnarray}
\left[\phi'_1+M_1+iM_2\right]\cos\theta&=&0,\nonumber \\
\left[\phi'_1-M_1-iM_2\right]\sin\theta&=&0\label{line_dLambda1}
\end{eqnarray}  
with
\begin{eqnarray}
& &M_1=2ge^{\sigma+2\phi_1}(\cosh\phi_0\cosh\phi_5\sinh\phi_2+\sinh\phi_0\sinh\phi_4\sinh\phi_5),\nonumber \\
& &M_2=-2ge^{\sigma+2\phi_1}\sinh\phi_0\cosh\phi_4\, .
\end{eqnarray}
For convenience, we also point out that the two equations in \eqref{line_dLambda1} arise from setting the coefficients of $(\chi^+\otimes \epsilon_0\otimes \eta,\chi^-\otimes \sigma_3\epsilon_0\otimes \eta)$ and $(\chi^+\otimes \sigma_3\epsilon_0\otimes \eta,\chi^-\otimes \epsilon_0\otimes \eta)$ to zero, respectively.
\\
\indent Consistency of these two equations implies $M_2=0$ which leads to 
\begin{equation}
\phi_0=0\, .
\end{equation}
Furthermore, to obtain non-trivial BPS equations for scalar fields, we have to set either $\sin\theta=0$ or $\cos\theta=0$ leading to the following BPS equations
\begin{equation}
\phi_1'=\mp M_1\, .
\end{equation}
This is exactly the same structure found in \cite{F4_defect} for compact gauging of the matter-coupled $F(4)$ gauged supergravity.
\\
\indent Similarly, $\delta\lambda^3_A$ and $\delta\lambda^4_A$ give
\begin{eqnarray}
\cosh\phi_5\phi'_2&=&\pm M_3,\nonumber \\
\cosh\phi_5\phi'_4&=&\pm \widetilde{M}_0,\nonumber \\
\phi_5'&=&\mp \widetilde{M}_3 \label{line_eq_1}
\end{eqnarray}
with
\begin{eqnarray}
& &M_3=2ge^\sigma \left[(e^{2\phi_1}-1)\cosh\phi_2-\sinh\phi_2\right],\nonumber \\
& &\widetilde{M}_0=2me^{-3\sigma}\cosh\phi_5\sinh\phi_4,\nonumber \\
& &\widetilde{M}_3=-2ge^\sigma\cosh\phi_5\sinh\phi_5+4ge^{\sigma+\phi_1}\sinh\phi_1\sinh\phi_2\sinh\phi_5\, .
\end{eqnarray}
In the three equations in \eqref{line_eq_1}, we have also used $\phi_0=0$ and $\sin\theta=0$ or $\cos\theta=0$. 
\\
\indent Consistency of the BPS equations from $\delta \chi_A$ and $\delta\psi_{A\mu}$ leads to another condition of the form
\begin{equation}
\sinh\phi_4\sinh\phi_5=0
\end{equation} 
which further implies that we can have solutions with either $\phi_4=0$ or $\phi_5=0$. However, it turns out that the two-form field equation requires $\phi_4=0$. Taking into account all of these results, we eventually arrive at the following set of the BPS equations
\begin{eqnarray}
\phi'_1&=&2ge^{\sigma+2\phi_1}\cosh\phi_5\sinh\phi_2,\\
\phi'_2&=&2ge^\sigma\textrm{sech}\phi_5\left[(e^{2\phi_1}-1)\cosh\phi_2-\sinh\phi_2\right],\\
\phi_5'&=&-2ge^\sigma\sinh\phi_5\left(\cosh\phi_2-2e^{\phi_1}\sinh\phi_1\sinh\phi_2\right),\\
\sigma'&=&\frac{1}{2}ge^\sigma\cosh\phi_2\cosh\phi_5-\frac{3}{2}me^{-3\sigma}-ge^{\sigma+\phi_1}\cosh\phi_5\sinh\phi_1\sinh\phi_2,\\
b'&=&2me^{-3\sigma}b,\\
f'&=&ge^{\sigma+\phi_1}\cosh\phi_5\sinh\phi_1\sinh\phi_2-\frac{1}{2}me^{-3\sigma}-\frac{1}{2}ge^\sigma\cosh\phi_2\cosh\phi_5,\\
Y'&=&\frac{1}{2}Yf',\qquad h'=f' 
\end{eqnarray}
together with two algebraic constraints
\begin{eqnarray}
b=-\frac{e^{f+\sigma}}{\ell m}\qquad \textrm{and}\qquad R=\frac{2\ell e^{2f-h+\sigma}}{2e^{f+\sigma}-\ell mb}\, .\label{line_defect_con}
\end{eqnarray}
In all the BPS equations given above, we have replaced the constant $R$ by using the second constraint given in \eqref{line_defect_con}. It is also useful to point out that the two algebraic constraints result from the fact that $\delta\chi_A$, $\delta \psi_{Aa}$ and $\delta\psi_{Ai}$ lead to three different BPS equations for $b(r)$. Taking one of them as a flow equation for $b(r)$ and requiring this equation be equivalent to the remaining two equations give rise to the aforementioned two algebraic constraints, see more detail in \cite{F4_defect}. In the BPS equations, we have also chosen a particular sign choice to identify the $AdS_6$ vacuum as $r\rightarrow \infty$.
\\
\indent Since the BPS equations for $h'$ and $f'$ are the same, we immediately find that $f=h+C$ for a constant $C$. The constant $C$ can be chosen to be zero by requiring that the solution becomes locally $AdS_6$ geometry with $f\sim h\sim \frac{r}{L}$. Therefore, we will set $f(r)=h(r)$ from now on. We have also verified that the BPS equations are compatible with the two algebraic constraints and the corresponding second-ordered field equations. We are not able to analytically solve these equations in full generality. This requires some numerical analysis which can always be achieved by suitable boundary conditions. We will instead consider some interesting solutions that can be obtained analytically within particular subtruncations. These solutions could be much more useful in most purposes. 
\\
\indent We first consider a subtruncation with $\phi_1=\phi_2=0$. This effectively turns off irrelevant operators dual to $\phi_1$ and $\phi_2$ at the $AdS_6$ vacuum. With a simpler set of equations, we can now solve for the following solution
\begin{eqnarray}
\phi_5&=&\ln\left[\frac{1+e^{-2g(\rho-\rho_0)}}{1-e^{-2g(\rho-\rho_0)}}\right],\nonumber \\
\sigma&=&\frac{1}{4}\ln\left[\frac{\sinh\phi_5(\sigma_0+3m\coth\phi_5)}{g}\right],\nonumber \\
f&=&\frac{1}{12}\ln(3m\cosh\phi_5+\sigma_0\sinh\phi_5)-\frac{1}{3}\ln\sinh\phi_5,\nonumber \\
b&=&-\frac{(\sigma_0+3m\coth\phi_5)^{\frac{1}{3}}}{\ell mg^{\frac{1}{4}}},\qquad Y=e^{\frac{f}{2}}
\end{eqnarray} 
with $\sigma_0$ being an integration constant and $\rho$ defined by $\frac{d\rho}{dr}=e^\sigma$. 
\\
\indent First of all, we notice that the solution approaches the $AdS_6$ vacuum with $\phi_5=0$ and $\sigma=\frac{1}{4}\ln\left[\frac{3m}{g}\right]$ for any value of $\sigma_0$. This solution takes the same form as the holographic RG flow studied in \cite{ISO3_flow} with a non-vanishing two-form field. This type of solutions is called charged domain walls in \cite{AdS2_F4_Dibitetto} and \cite{6D_surface_Dibitetto}. Near the $AdS_6$ vacuum, we find
\begin{eqnarray}
\phi_5\sim \sigma\sim e^{-\frac{3r}{L}}\qquad \textrm{and}\qquad b\sim \phi_5^{-\frac{1}{3}}\sim e^{\frac{r}{L}} 
\end{eqnarray}  
in which we have set $g=3m$. From this behavior, we see that $\phi_5$ and $\sigma$ correspond to vacuum expectation values of relevant operators of dimension $3$ while $b$ is dual to a dimension-$4$ operator as in the solutions considered in \cite{line_defect_F4_Gutperle,F4_defect}. The asymptotic behavior also indicates that the line defect involves a source of this dimension-$4$ operator. The solution is singular at $\rho=\rho_0$ with
\begin{equation}
\phi_5\sim -\ln(\rho-\rho_0)
\end{equation}
and
\begin{eqnarray}
& &\sigma \sim \frac{1}{4}\ln(\rho-\rho_0),\qquad f\sim \frac{5}{12}\ln(\rho-\rho_0),\qquad b\sim -\frac{6^{\frac{1}{3}}(\rho-\rho_0)^{\frac{2}{3}}}{\ell g^{\frac{1}{4}}m^{\frac{2}{3}}} 
\end{eqnarray}
for $\sigma_0=-3m$, and
\begin{eqnarray}
& &\sigma\sim -\frac{1}{4}\ln(\rho-\rho_0),\qquad f\sim \frac{1}{4}\ln(\rho-\rho_0),\qquad b\sim -\frac{(3m+\sigma_0)^{\frac{1}{3}}}{\ell g^{\frac{1}{4}}m} 
\end{eqnarray}
for $\sigma_0\neq -3m$. These are the same asymptotic behaviors of scalars studied in \cite{ISO3_flow} in the case of holographic RG flows to non-conformal phases of the $N=2$ SCFT.
\\
\indent Another solution we will consider is given by $\phi_5=0$ and $\phi_2=\phi_1$. Unlike the previous solution with $SO(2)\times U(1)$ symmetry, this solution preserves a larger $SO(3)\times U(1)$ symmetry and is given by
\begin{eqnarray}
g(\rho-\rho_0)&=&e^{-\phi_1}+\frac{1}{2}\ln\left[\frac{1-e^{-\phi_1}}{1+e^{-\phi_1}}\right],\nonumber \\
\sigma&=&\frac{1}{4}\ln\left[\frac{e^{-\phi_1}(\tilde{\sigma}_0e^{4\phi_1}-3m)}{2g(e^{2\phi_1}-1)}\right],\nonumber \\
f&=&\frac{1}{4}\ln(1-e^{2\phi_1})+\frac{1}{12}\ln(3m-\tilde{\sigma}_0e^{4\phi_1})-\frac{13}{12}\phi_1,\nonumber \\
b&=&-\frac{e^{-\frac{4}{3}\phi_1}(3m-\tilde{\sigma}_0e^{4\phi_1})^{\frac{1}{3}}}{\ell m(2g)^{\frac{1}{4}}},\qquad Y=e^{\frac{f}{2}}
\end{eqnarray}
with the same coordinate $\rho$ as in the previous case and another integration constant $\tilde{\sigma}_0$. This solution also takes the form of a charged domain wall. In this case, however, we need to set $\tilde{\sigma}_0=3m$ in order to obtain the asymptotically $AdS_6$ geometry as $r\rightarrow \infty$. Near the $AdS_6$ vacuum, we find
\begin{equation}
\phi_1\sim e^{\frac{3r}{L}},\qquad \sigma\sim \phi_1^{-1}\sim e^{-\frac{3r}{L}},\qquad b\sim \phi_1^{\frac{1}{3}}\sim e^{\frac{r}{L}}\, .
\end{equation}
As in the previous case, there is a source for an operator of dimension $4$ dual to the two form field $b$. The dilaton $\sigma$ and $\phi_1$ correspond to a vacuum expectation value of a dimension-$3$ operator as before and a source of a dimension-$8$ operator, respectively. As $\rho\rightarrow \rho_0$, we find, see more detail in \cite{ISO3_flow},
\begin{eqnarray}
& &\phi_1\sim -\frac{1}{3}\ln(\rho-\rho_0),\qquad \sigma\sim -\frac{1}{12}\ln(\rho-\rho_0),\nonumber \\
& &f\sim \frac{1}{12}\ln(\rho-\rho_0),\qquad b\sim -\frac{3^{\frac{1}{3}}}{(2g)^{\frac{1}{4}}m^{\frac{2}{3}}\ell}\, .
\end{eqnarray}
This behavior is again the same as the RG flow solution considered in \cite{ISO3_flow} with a non-vanishing two-form field.

\subsection{Surface defects}
We now repeat a similar analysis for solutions describing conformal surface defects within the $N=2$ SCFT in five dimensions. The metric ansatz is now given by
\begin{equation}
ds^2=e^{2f(r)}dx^2_{AdS_3}+dr^2+e^{2h(r)}ds^2_{S^2}\, .\label{surface_metric}
\end{equation}
The solutions preserve two-dimensional conformal symmetry corresponding to the $AdS_3$ factor. Similar to the previous analysis, we will also turn on the two-form field with the ansatz
\begin{equation}
B=b(r)\textrm{vol}_{S^2}\, .
\end{equation}
The scalar fields are given by the coset representative in \eqref{L_SO2}.
\\
\indent In this case, we will split the six-dimensional coordinates as $x^\mu=(x^a,r,\theta^i)$ with $a=0,1,2$ and $i=1,2$ and take the Killing spinors to be 
\begin{equation}
\epsilon=Y\eta\otimes\left[\cos\theta\chi^+\otimes \epsilon_0+i\sin\theta\gamma_*\chi^+\otimes \sigma_3\epsilon_0-i\cos\theta\gamma_*\chi^-\otimes \sigma_3\epsilon_0+\sin\theta\chi^-\epsilon_0\right].
\end{equation}
The Killing spinors on $AdS_3$ and $S^2$, denoted by $\eta$ and $\chi^\pm$, respectively satisfy
\begin{equation}
\hat{\nabla}_a=\frac{1}{2\ell}\rho_a \eta\qquad \textrm{and}\qquad \tilde{\nabla}_i\chi^\pm=\frac{1}{2R}\gamma_*\tilde{\gamma}_i\chi^\mp\, .\label{Killing_eq_surface}
\end{equation}
In this case, $\ell$ and $R$ are respectively $AdS_3$ and $S^2$ radii. The gamma matrices are chosen to be
\begin{eqnarray}
& &\gamma^a=\rho^a\otimes \mathbb{I}_2\otimes \sigma,\qquad \gamma^r=-(\mathbb{I}_2\otimes \gamma_*\otimes \sigma_1),\nonumber \\
& &\gamma^{i}=-(\mathbb{I}_2\otimes \tilde{\gamma}^i\otimes \sigma_1), \qquad \gamma_7=i\mathbb{I}_2\otimes \mathbb{I}_2\otimes \sigma_3
\end{eqnarray}
with $\gamma_*=i\tilde{\gamma}^1\tilde{\gamma}^2$ and $\gamma_*^2=\mathbb{I}_2$.
\\
\indent To derive the BPS equations, we impose the following projector on the Killing spinors 
\begin{equation}
(\gamma_*\otimes \sigma_1)(\chi^\pm\otimes \epsilon_0)=\pm \chi^\pm \otimes \epsilon_0\, .
\end{equation}
It is also useful to note that this projector implies 
\begin{equation}
\chi^\pm\otimes \sigma_2\epsilon_0=\mp i\gamma_* \chi^\pm \otimes \sigma_3\epsilon_0\, .
\end{equation}
The same analysis as in the case of line defects shows that the gaugino variations restrict the phase of the Killing spinors to satisfy $\sin\theta=0$ or $\cos\theta=0$. Consistency among the BPS equations again imposes the conditions $\phi_0=\phi_4=0$ in addition to $\phi_3=0$ required by a consistent truncation of gauge fields.
\\
\indent With all these and an appropriate sign choice to identify the $AdS_6$ vacuum with $r\rightarrow \infty$, we end up with the following set of BPS equations
\begin{eqnarray}
\phi'_1&=&2ge^{\sigma+2\phi_1}\cosh\phi_5\sinh\phi_2,\\
\phi_2'&=&2ge^\sigma\textrm{sech}\phi_5\left[(e^{2\phi_1}-1)\cosh\phi_2-\sinh\phi_2\right],\\
\phi_5'&=&-2ge^\sigma\sinh\phi_5(\cosh\phi_2-2e^{\phi_1}\sinh\phi_1\sinh\phi_2),\\
\sigma'&=&\frac{3}{2}me^{-3\sigma}-\frac{1}{2}g\cosh\phi_5(\cosh\phi_2-2e^{\phi_1}\sinh\phi_1\sinh\phi_2),\\
f'&=&\frac{1}{2}ge^\sigma\cosh\phi_5(\cosh\phi_2-2e^{\phi_1}\sinh\phi_1\sinh\phi_2)+\frac{1}{2}me^{-3\sigma},\\
Y'&=&\frac{1}{2}Yf', \qquad h'=f',\\
b'&=&2me^{-3\sigma}b
\end{eqnarray}
together with two algebraic constraints
\begin{equation}
b=-\frac{2e^{2h-f+\sigma}}{\ell m}\qquad \textrm{and}\qquad R=\frac{2\ell e^{f+h+\sigma}}{2e^{2h+\sigma}-\ell mbe^f}\, .\label{surface_defect_con}
\end{equation}
\indent As in the case of line defects, we have replaced the constant $R$ in the BPS equations by using the second constraint given in \eqref{surface_defect_con}. We now consider two types of solutions that can be analytically obtained in particular subtruncations. The first type of solutions is found by setting $\phi_2=\phi_1=0$ which leads to the following solution
\begin{eqnarray}
\phi_5&=&\ln\left[\frac{1+e^{-2g(\rho-\rho_0)}}{1-e^{-2g(\rho-\rho_0)}}\right],\nonumber \\
\sigma&=&\frac{1}{4}\ln\left[\frac{\sinh\phi_5(\sigma_0+3m\coth\phi_5)}{g}\right],\nonumber \\
f&=&\frac{1}{12}\ln(3m\cosh\phi_5+\sigma_0\sinh\phi_5)-\frac{1}{3}\ln\sinh\phi_5,\nonumber \\
b&=&-\frac{2}{\ell mg^{\frac{1}{4}}}(\sigma_0+3m\coth\phi_5)^{\frac{1}{3}} ,\qquad Y=e^{\frac{f}{2}}
\end{eqnarray}
where, as before, $\rho$ is defined by $\frac{d\rho}{dr}=e^\sigma$ and $h=f$. 
\\
\indent Another type of solutions is obtained by setting $\phi_5=0$ and $\phi_2=\phi_1$ giving rise to the solution
\begin{eqnarray}
g(\rho-\rho_0)&=&e^{-\phi_1}+\frac{1}{2}\ln\left[\frac{1-e^{-\phi_1}}{1+e^{-\phi_1}}\right],\nonumber \\
\sigma&=&\frac{1}{4}\ln\left[\frac{e^{-\phi_1}(\tilde{\sigma}_0e^{4\phi_1}-3m)}{2g(e^{2\phi_1}-1)}\right],\nonumber \\
f&=&\frac{1}{4}\ln(1-e^{2\phi_1})+\frac{1}{12}\ln(3m-\tilde{\sigma}_0e^{4\phi_1})-\frac{13}{12}\phi_1,\nonumber \\
b&=&-\frac{2^{\frac{3}{4}}e^{-\frac{4}{3}\phi_1}(3m-\tilde{\sigma}_0e^{4\phi_1})^{\frac{1}{3}}}{\ell mg^{\frac{1}{4}}} ,\qquad Y=e^{\frac{f}{2}}\, .
\end{eqnarray}
with the same coordinate $\rho$ as in the previous solution. The integration constant $\tilde{\sigma}_0$ needs to be chosen as $\tilde{\sigma}_0=3m$ in order for the solution to approach the $AdS_6$ vacuum. All of these solutions take the same form as the solutions for line defects considered above with the only difference being a numerical factor in the solution for $b$. Asymptotic behaviors are the same as in the case of line defects, and we will not repeat these again. We simply point out that the surface defects also arise from turning on a dimension-$4$ operator dual to the two-form field as in the case of line defects. Although the solutions in both cases take the same form, we expect that the two types of solutions lead to different uplifted ten-dimensional solutions. In particular, in the case of line defects, we have an electric component of the two-form field while the magnetic component appears in the case of surface defects. 
\\
\indent We end this section by giving some comments regarding the solutions of line and surface defects in matter-coupled $F(4)$ gauged supergravity with a compact $SU(2)\times U(1)$ gauge group considered in \cite{F4_defect} and a non-semisimple $ISO(3)\times U(1)$ gauge group studied in this paper. We note that although the $SO(2)\times U(1)$ symmetric solutions given here take the same from as those in \cite{F4_defect}, the solutions presented in this paper are obtained within a particular subtruncation with $\phi_1=\phi_2=0$. The more general defect solutions, obtained by including these two scalars dual to irrelevant operators in the five-dimensional $N=2$ SCFT, are certainly different from the solutions found in \cite{F4_defect}. The latter are, on the other hand, the most general solutions within the $SO(2)\times U(1)$ symmetric sector not in any particular subtruncations. Therefore, the crucial difference between the solutions from compact and non-semisimple gauge groups lies in the presence of the dual irrelevant operators. 
%%%%%%%%%%%%%%%%%%%%%%%%%%%%%%%%%%%%%%%%%%%%%%%%%%%%%%%%%%%%%%%%%%
\section{Conclusions}\label{conclusion}
We have studied a number of supersymmetric solutions from matter-coupled $F(4)$ gauged supergravity with $ISO(3)\times U(1)$ gauge group. The first class of these solutions takes the form of RG flows across dimensions and holographically describes twisted compactifications of the five-dimensional $N=2$ SCFT on $H^2$ and $H^3$ to three- and two-dimensional SCFTs in the IR. These solutions have been obtained by performing standard topological twists using $SO(2)\times U(1)$ and $SO(3)$ gauge fields. Since the $ISO(3)\times U(1)$ $F(4)$ gauged supergravity is a consistent truncation of type IIB theory on $S^2\times \Sigma$, the $AdS_4\times H^2$ and $AdS_3\times H^3$ solutions found in this paper could lead to new supersymmetric $AdS_4\times H^2\times S^2\times \Sigma$ and $AdS_3\times H^3\times S^2\times \Sigma$ solutions in type IIB theory.  
\\
\indent The second class of solutions obtained in this paper takes the form of ``charged'' curved domain walls with $AdS_2\times S^3$ and $AdS_3\times S^2$ slicing. These solutions have been found by turning on the two-form field in the gravity multiplet to support the curvature of the domain walls. The solutions have a holographic interpretation as conformal line and surface defects within the $N=2$ SCFT in five dimensions. These defects arise from a source term for a dimension-$4$ operator dual to the two-form field. The solutions also have a very similar structure as those found in compact gauge groups studied in \cite{F4_defect}. In particular, the solutions are mainly holographic RG flows in the presence of a running two-form field. 
\\
\indent Although the matter-coupled $ISO(3)\times U(1)$ $F(4)$ gauged supergravity considered here can be embedded in type IIB theory, the complete truncation ansatz has not been constructed to date. It would be interesting to work this out using the formalism of exceptional field theories and use the resulting truncation ansatz to uplift the solutions found in this paper to ten dimensions. This could lead to new $AdS_4\times H^2$ and $AdS_3\times H^3$ solutions in type IIB theory as mentioned above as well as possible brane configurations with near horizon geometries given by these solutions. It would also be interesting to identify the dual $N=2$ SCFT in five dimensions together with field theory descriptions of conformal line and surface defects dual to the solutions found here in terms of position-dependent deformations. 
\vspace{0.5cm}\\
%%%%%%%%%%%%%%%%%%%%%%%%%%%%%%%%%%%%%%%%%%%%%%%%%%%%%%%%%%%%%
{\large{\textbf{Acknowledgement}}} \\
This work is funded by National Research Council of Thailand (NRCT) and Chulalongkorn University under grant N42A650263.  
%%%%%%%%%%%%%%%%%%%%%%%%%%%%%%%%%%%%%%%%%%%%%%%%%%%%%%%%

\end{document}